\documentclass[fleqn,usenatbib]{mnras}

\usepackage{newtxtext,newtxmath}
\usepackage[T1]{fontenc}
\DeclareRobustCommand{\VAN}[3]{#2}
\let\VANthebibliography\thebibliography
\def\thebibliography{\DeclareRobustCommand{\VAN}[3]{##3}\VANthebibliography}
\usepackage{graphicx}
\usepackage{amsmath}
\usepackage[normalem]{ulem}
\usepackage{xcolor}

\title[Nucleosynthesis in 3D AIC models]{Nucleosynthesis in the Accretion-induced Collapse of Rotating White Dwarfs and Impact of Magnetic Fields}

\author[L. Th\"ummler, T. Kuroda, and M. Shibata]{
Laurenz Th\"ummler$^{1}$\thanks{E-mail: lthuemmler@ethz.ch},
Takami Kuroda$^{2}$\thanks{E-mail: takami.kuroda@aei.mpg.de},
and Masaru Shibata$^{2,3}$
\\
$^{1}$Department of Mathematics, ETH Z\"urich, R\"amistrasse 101, 8092 Z\"urich, Switzerland\\
$^{2}$Max-Planck-Institut f{\"u}r Gravitationsphysik, Am M{\"u}hlenberg 1, D-14476 Potsdam-Golm, Germany\\
$^{3}$Center for Gravitational Physics and Quantum-Information,
Yukawa Institute for Theoretical Physics, Kyoto University, Kyoto, 606-8502, Japan
}

\date{Accepted XXX. Received YYY; in original form ZZZ}
\pubyear{2026}

\begin{document}
\label{firstpage}
\pagerange{\pageref{firstpage}--\pageref{lastpage}}
\maketitle

\begin{abstract}
Accretion-induced collapse (AIC) of a rotating oxygen-neon-magnesium white dwarf (WD) is an alternative channel of neutron-star formation, in which an accreting WD collapses instead of exploding thermonuclearly.
If sufficiently magnetised, the progenitor can drive magnetorotational outflows analogous to those of magnetorotational supernovae, and may be a site for rapid neutron-capture ($r$-process) nucleosynthesis.
We post-process tracer particles from five three-dimensional general-relativistic neutrino-(magneto)hydrodynamic AIC simulations with the WinNet nuclear reaction network: four non-magnetised models spanning four initial rotation rates and a magnetised counterpart of the fastest rotator.
Ejecta mass, the fraction of matter reaching nuclear statistical equilibrium, and the ejected mass beyond the iron group all increase monotonically with rotation rate, with magnetic fields further amplifying these trends.
The slowest-rotating model synthesises essentially only iron-group material, while the most rapidly rotating and magnetised model reaches the second $r$-process peak at mass number $A\approx130$, with a trace, non-robust signal in the actinide region (up to $A\approx244$).
The corresponding $^{56}$Ni mass rises from $0.007$ to $0.014\,{\rm M}_\odot$, five to over forty times below representative core-collapse supernova and hypernova yields, implying a faint radioactively powered transient.
Contrary to the picture for magnetorotationally driven supernovae, the heaviest ejecta do not track the polar outflow: they reside in equatorial-to-mid-latitude, moderate-entropy lobes, while the polar column is a local minimum in some heavy-element abundances.
We conclude that rapidly rotating, magnetised AIC events could contribute to the Galactic yields of trans-iron and weak $r$-process nuclei, although no model produces a robust third $r$-process peak.
\end{abstract}


\section{Introduction}
\label{sec:intro}

A white dwarf that accretes mass towards the Chandrasekhar limit can meet one of two
qualitatively different fates depending on its composition and accretion rate: a thermonuclear runaway that disrupts the star (the progenitor channel for Type~Ia supernovae), or electron-capture reactions on $^{24}$Mg/$^{24}$Na and $^{20}$Ne, that reduce the electron fraction and heat the degenerate ONe(Mg) core.
The latter reactions can trigger oxygen ignition and a deflagration; at sufficiently high density, continued neutronisation of the burned material favours accretion-induced collapse (AIC) rather than thermonuclear disruption \citep{Miyaji1980,NomotoKondo1991,Schwab2015,Jones2016}.
In this case, the progenitor WD is not destroyed completely, but leaves behind a neutron star (NS).

Binary-population-synthesis calculations predict that AIC can arise through several distinct pathways, including stable accretion on to an ONe white dwarf from a main-sequence, red-giant, or helium-star donor, as well as merger-induced collapse in double-degenerate systems \citep{Wang2018,Ruiter2019,Tauris2013,Wang2020b}.
The corresponding galactic event rate remains uncertain because it depends on the treatment of common-envelope evolution, mass-accretion history, off-centre carbon ignition, all of which bifurcate the fate into the thermonuclear disruption and collapse branches \citep{NomotoKondo1991,Schwab2015,WuWang2018,Wang2018}.
Population-synthesis calculations nevertheless generally place AIC below the core-collapse supernova rate, while allowing it to constitute a small but potentially astrophysically significant population of NS-forming events \citep{Fryer99,Ruiter2019,Wang2020b}.

Their importance for galactic evolution, however, may extend beyond the NS formation channel.
The neutron-rich ejecta from AIC events have long been investigated as a potential source of rare neutron-rich isotopes and neutron-capture elements, and could therefore contribute to galactic chemical evolution \citep{Woosley1992,Fryer99}.
Indeed, a recent one-dimensional, non-magnetised AIC calculation found substantial production of first-peak neutron-capture elements but only a small $^{56}$Ni yield \citep{Yip2024}.
Long-duration axisymmetric full-fledged neutrino-radiation hydrodynamics simulations further showed that rotation can qualitatively alter the electron-fraction evolution of the ejecta, producing increasingly neutron-rich bipolar outflows at late times with the potential for $r$-process nucleosynthesis \citep{Batziou2025}.
Moreover, axisymmetric general-relativistic neutrino-magnetohydrodynamics (MHD) simulations of rapidly rotating, strongly magnetised WDs could produce relativistic jets and neutron-rich outflows that could support heavy $r$-process nucleosynthesis and power kilonova emission \citep{Cheong25_AIC}.
Building on those models, \citet{Pitik2026} post-processed an axisymmetric magnetised AIC model with a nuclear reaction network and predicted the resulting kilonova-like emission, so that the influence of magnetic fields on AIC nucleosynthesis and on its electromagnetic counterpart has already been quantified in axisymmetry.
The sensitivity of these detailed AIC yields to rotation, magnetic field, and genuinely three-dimensional dynamics including ejecta geometry, however, remains largely unexplored.

Whether this NS formation channel and the associated ejecta composition make an appreciable contribution to galaxies and their chemical evolution depends on the AIC occurrence rate.
Binary population-synthesis calculations predict a total Galactic AIC rate of approximately $(1.7$--$9.8)\times10^{-3}\ {\rm yr}^{-1}$ when both single- and double-degenerate channels, including double-CO-WD mergers, are counted.
Excluding the uncertain double-CO-WD channel reduces the predicted rate to $(0.6$--$4.7)\times10^{-3}\ {\rm yr}^{-1}$ \citep{Liu2020,Wang2020b}.
For a consistent Galactic comparison, the star-formation rate of $5\,{\rm M}_\odot\,{\rm yr}^{-1}$ adopted by \citet{Liu2020} corresponds to a model-implied Galactic core-collapse supernova rate of approximately $4.6\times10^{-2}\ {\rm yr}^{-1}$.
The predicted AIC rate is therefore approximately $4$--$21$~per cent of the massive-star core-collapse supernova rate when all channels are included, or approximately $1$--$10$~per cent when the double-CO-WD channel is excluded.
Although these estimates remain highly model dependent, they suggest that the cumulative contribution of AIC ejecta to Galactic chemical evolution may be non-negligible.

The nucleosynthetic outcome of an individual AIC depends sensitively on the progenitor properties, which can strongly affect both the amount and composition of the ejecta; among these properties, the rotation of the WD is expected to play a particularly important role.
From the perspective of progenitor WD evolution theory, early evolutionary calculations highlighted angular-momentum transfer from the companion as a primary mechanism for spinning up the WD.
Angular-momentum accretion can substantially modify both the equilibrium structure and, consequently, the evolution towards the eventual unstable WD configuration \citep{Piersanti2003,Uenishi2003,Saio2004,YoonLanger2005,Liu2001}.
More recent binary- and stellar-evolution calculations have refined the accretion-rate and donor-star parameter space leading to collapse rather than thermonuclear disruption \citep{WuWang2018,Wang2020b}.

AIC simulations have correspondingly shown that rotation strongly affects the collapse dynamics and outflow geometry.
Due to essentially the complete absence of a progenitor WD envelope, the collapse is followed by a prompt-type explosion, dominated mainly by a bipolar structure \citep{Fryer99,Dessart2006,Abdikamalov2010,LongoMicchi2023,Batziou2025,Kuroda2025}, with an ejecta mass significantly smaller than that of typical core-collapse supernovae.
Modern multi-dimensional AIC simulations comprise axisymmetric radiation-hydrodynamic \citep{Dessart2006,Batziou2025} and general-relativistic \citep{Abdikamalov2010} models, axisymmetric magnetohydrodynamic models \citep{Dessart2007}, axisymmetric general-relativistic neutrino-magnetohydrodynamic models \citep{Cheong25_AIC,Pitik2026,LongoMicchi2026}, and fully 3D general-relativistic models without \citep{LongoMicchi2023,Kuroda2025} and with magnetic fields \citep{Kuroda2025,Combi2025}.
Furthermore, the ejecta maintain nearly their original high velocities, exhibiting little to no significant deceleration throughout their expansion till they eventually collide either with the common envelope of binary WDs or with the envelope of a massive companion star \citep{Schwab21}.

The latest full three-dimensional (3D) general relativistic (GR) studies \citep{LongoMicchi2023,Kuroda2025} reported the significance of 3D effects, when the post-collapse remnant is strongly and differentially rotating.
These numerical studies have identified one-armed, low-$T/|W|$ shear instabilities that can grow below the classical dynamical bar-mode threshold \citep{Shibata2002,Centrella2001,Watts2005,Corvino2010}.
Such rotation-induced non-axisymmetric instabilities can produce strong gravitational-wave emission \citep{LongoMicchi2023,Kuroda2025} and facilitate the efficient ejection of neutron-rich matter from the vicinity of the nascent neutron-star surface \citep{Kuroda2025}. Fully 3D AIC simulations are therefore essential not only for understanding the explosion dynamics but also for predicting nucleosynthetic yields and the broader multi-messenger signatures of these events, including gravitational waves and electromagnetic (EM) emission powered by the radioactive decay of freshly synthesised isotopes.

The AIC of a magnetised WD is a natural extension of this picture.
From a viewpoint of plausible formation process of magnetised WD, the merger of two WDs may amplify initially weak seed magnetic fields to dynamically significant strengths \citep{Tout08,Garcia-Berro12}, as found in magnetohydrodynamic simulations of WD mergers \citep{Ji2013,Pakmor2024}.
Indeed, the existence of such strongly magnetised WDs is observationally supported by \citet{Wickramasinghe00,Schmidt03} and by more recent reviews and volume-limited surveys \citep{Ferrario2015,Bagnulo2021,Bagnulo2022}, which find strong fields to be particularly common in high-mass WDs that are probably merger products; a highly magnetised, rapidly rotating WD with a mass close to the Chandrasekhar mass has also been observed \citep{Caiazzo2021}, as well as by numerical simulations of WD mergers in \citet{Zhu15}.
These fields can be further amplified by flux compression during collapse and subsequently by rotational winding and nonlinear magnetohydrodynamic processes, including the magnetorotational instability \citep{Balbus91}, the Tayler--Spruit dynamo \citep{Spruit02,Reboul-Salze24}, and convection-driven turbulent dynamos operating within the nascent neutron star \citep{ThompsonDuncan1993,Raynaud2020,Masada2022}.
These mechanisms are not specific to AIC but may operate more generally in astrophysical systems in which the requisite rotational, magnetic, and thermodynamic conditions are realised.
Consequently, some AIC events may produce poloidal magnetic fields of order $B_{\rm p}\sim10^{15}\,{\rm G}$, providing a viable formation channel for rapidly rotating magnetars and resultant magnetorotationally driven outflows as has been demonstrated by \citet{Dessart2007,Cheong25_AIC,Kuroda2025,LongoMicchi2026}.
The global magnetic field structure can modify the polar and equatorial mass fluxes and may therefore be directly relevant to the rotation-dependent angular abundance structure and, consequently, the observational feature of EM counterparts of these kinds, as shown in axisymmetry by \citet{Pitik2026}.

These magnetorotational effects are particularly
important for nucleosynthesis because they alter the expansion time, entropy, neutrino
exposure, and electron fraction of the unbound matter.
 The conditions reached in the innermost, most neutron-rich AIC outflows (high entropy, low electron fraction $Y_e$, and short expansion time-scale) are naturally expected to be analogous to those invoked for rapid neutron-capture ($r$-process) nucleosynthesis in jet-driven magnetorotational supernovae (MR-SNe) of massive stars
\citep{Winteler2012,Nishimura2015,Nishimura2017,Halevi2018,Mosta2018}, including recent 3D models \citep{Zha2024}.
The connection between magnetic-field amplification, non-axisymmetric dynamics, and
neutron-rich jet ejecta has been developed through progressively more realistic multi-dimensional
simulations~\citep{Nishimura2006,Mosta2014,Mosta2015,Reichert2021}, more recently also with systematic variations of the magnetic field \citep{ObergaulingerAloy2020,Reichert2024} and in 3D \citep{Powell2023}.
This has been quantified in axisymmetry \citep{Pitik2026}, but whether rotating AIC can likewise act as an $r$-process site, and how strongly this depends on rotation rate, has not been addressed with detailed nuclear-reaction-network post-processing of a 3D AIC simulation, in which the non-axisymmetric dynamics discussed above are captured.

Here we post-process tracer particles from the 3D general-relativistic neutrino-radiation-(magneto)hydrodynamics AIC simulations of \citet{Kuroda2025} with the nuclear reaction network WinNet \citep{Reichert2023}, following the same tracer-particle-plus-network methodology originally used to establish the MR-SNe jet $r$-process \citep{Winteler2012} and applied in detail to MR-SNe by \citet{Reichert2023b}.
We adopt the figure set and analysis approach of \citet{Reichert2023b}, namely ejected mass and mass-fraction distributions, overproduction factors relative to solar abundances, electron-fraction and hot/cold-ejecta statistics, spatial (meridional) maps of entropy and heavy-element content, and $^{56}$Ni yield, but apply it across a sequence of four initial rotation rates of the \emph{same} AIC progenitor family, complemented by a magnetised counterpart of the fastest-rotating model, rather than to a single magnetorotational core-collapse model, isolating rotation (and, incidentally, explosion energetics) as the controlling parameter.
We provide the corresponding 3D, short-term ($\lesssim1\,{\rm s}$) nucleosynthetic yields and their spatial distribution.

The paper is organised as follows. Section~\ref{sec:methods} describes the hydrodynamical models, the
tracer-particle and reaction-network post-processing methodology, and the solar reference abundances.
Section~\ref{sec:results} presents the ejected mass and composition, overproduction factors, electron
fraction and hot/cold ejecta statistics, radioactive-isotope proxies, and the spatial distribution of
entropy and heavy elements. Section~\ref{sec:discussion} discusses these results in the context of the
jet $r$-process mechanism and the closest comparable studies, and Section~\ref{sec:conclusions}
summarises our conclusions.
Throughout this paper, cgs units are used.

\section{Numerical methods}
\label{sec:methods}
In this section, we first present all hydrodynamical models used in the nucleosynthesis calculations---four previously published models and one newly developed model, which is summarised in Appendix~\ref{app:R6B11}. We then briefly describe the nucleosynthesis code employed, along with the nuclear reaction networks and observational datasets used for comparison with the simulation results.

\subsection{AIC models}
\label{sec:models}

The hydrodynamical models post-processed here are taken from the 3D general-relativistic
neutrino-radiation-(magneto)hydrodynamics simulations of rotating, accretion-induced collapse of WDs presented by \citet{Kuroda2025}, which solve the Einstein equations in the BSSN/Z4c formalism with a two-moment (M1), multi-energy-bin, three-flavour neutrino transport scheme \citep{Shibata2011,Kuroda2016,Kuroda2020}.
For the nuclear equation of state (EOS), we adopt the SFHo relativistic mean-field EOS \citep{Hempel2010,SFH} (extended to lower densities as described in \citealt{Kuroda2025}).

Initial WDs are constructed in hydrostatic equilibrium for a fixed
central density $\rho_{\rm c,0}=10^{10}\,{\rm g\,cm^{-3}}$, uniform (rigid) angular frequency $\Omega_0$, electron fraction
$Y_e=0.5$ and entropy $s=0.8\,k_{\rm B}\,{\rm baryon^{-1}}$, parametrised by the ratio of rotational kinetic to gravitational potential energy $\beta_0 \equiv |T_{\rm rot}/W|$.
Here we define the rotational kinetic and gravitational potential energy as \citep{Friedman86}
\begin{align}
T_{\rm rot} &\equiv \frac{1}{2}\int d^3x\, \rho \sqrt{\gamma}
    W_{\rm L} u_\varphi \Omega, \label{eq:Trot}\\
W &\equiv M_{\rm bar}+E_{\rm int}+T_{\rm rot}-M_{\rm ADM},\label{eq:W}
\end{align}
with $\rho$, $\gamma$, $W_{\rm L}$, $u^\varphi({\rm or}~u_\phi)$, $\Omega$, $M_{\rm bar}$, $E_{\rm int}$, $M_{\rm ADM}$, being the rest mass density, determinant of three metric, Lorentz factor, azimuthal component of four-velocity, local angular frequency (i.e., initially $\Omega=\Omega_0$ throughout the current rigidly rotating WD), baryonic mass of WD, internal energy of WD, Arnowitt--Deser--Misner (ADM) mass of the system, respectively.
We use four non-magnetized models spanning the rotation-rate sequence, adopting the same R1, R2, R3 and R6 notation as \citet{Kuroda2025} (for these models $\beta_0\approx0.1,\,0.2,\,0.3,\,0.6\,{\rm per\,cent}$, respectively).
These four models form the basis for the discussion in the current study.
Our fifth, newly developed model R6B11 shares the same initial rotation ($\beta_0$, $J_0$, $\Omega_0$) as R6 but is a magnetised counterpart of that model.
The initial magnetic field configuration is identical to that of our previous octant symmetry model R6oB in \citet{Kuroda2025}, except we do not enforce any octant symmetry, and the central magnetic field strength $B_0(=10^{11}$\,G) and the radius of the central sphere with a uniform magnetic field $R_0(=10^{8}$\,cm) are adopted.

\begin{table*}
	\centering
	\caption{Initial parameters and explosion diagnostics of the five models analysed in this work,
	following the definitions of \citet{Kuroda2025}. $\beta_0$: initial ratio of rotational kinetic to gravitational potential energy, $|T_{\rm rot}/W|$ (Eqs.~\ref{eq:Trot} and~\ref{eq:W}); $\rho_{\rm c,0}$: initial central density; $L_{\rm box}$: size of the computational domain; $J_0$ and $\Omega_0$: initial angular momentum and angular velocity; $M_{\rm WD,bar}$: baryonic mass of the initial WD; $\beta_{\rm cb}$: the same energy ratio evaluated at core bounce; $t_9$: post-bounce time at which the maximum shock radius first reaches $10^{9}$\,cm; $E_{\rm exp,9}$ and $E_{\rm exp,fin}$: diagnostic explosion energy at $t_9$ and at $t_{\rm fin}$; $t_{\rm fin}$: post-bounce time at which the simulation is stopped. $*$: $t_{\rm fin}$ and $E_{\rm exp,fin}$ computed directly from each
	simulation's global-dynamics time series; R6B11 is our newly developed model and does not appear in \citealt{Kuroda2025}, this additionally applies to $t_9$ and $E_{\rm exp,9}$. See section~\ref{sec:models}.}
	\label{tab:models}
	\begin{tabular}{lccccccccccc}
		\hline
		Model & $\beta_0$ & $\rho_{\rm c,0}$ & $L_{\rm box}$ & $J_0$ & $M_{\rm WD,bar}$ & $\Omega_0$ & $\beta_{\rm cb}$ & $t_9$ & $E_{\rm exp,9}$ & $t_{\rm fin}$ & $E_{\rm exp,fin}$ \\
		 & (per cent) & (g\,cm$^{-3}$) & ($10^9$\,cm) &($10^{49}$\,g\,cm$^2$\,s$^{-1}$) & (M$_\odot$) & (rad\,s$^{-1}$) & (per cent) & (ms) & ($10^{50}$\,erg) & (ms) & ($10^{50}$\,erg) \\
		\hline
		R1 & 0.113 & $10^{10}$ & $12.0$& 1.05 & 1.73 & 1.38 & 0.56 & 338 & $1.03$ & $617^*$ & $1.16^*$ \\
		R2 & 0.217 & $10^{10}$ &$1.5$& 2.08 & 1.74 & 1.91 & 1.09 & 302 & 2.54 & $418^*$ & $2.79^*$ \\
		R3 & 0.320 & $10^{10}$ &$1.5$& 2.98 & 1.75 & 2.31 & 1.35 & 317 & 3.37 & $445^*$ & $3.74^*$ \\
		R6 & 0.647 & $10^{10}$ &$12.0$& 5.86 & 1.76 & 3.24 & 3.56 & 345 & 3.11 & $508^*$ & $4.27^*$ \\
		R6B11 & 0.647 & $10^{10}$ & $12.0$&5.86 & 1.76 & 3.24 & 3.56 & $335^*$ & $6.01^*$ & $764^*$ & $8.52^*$ \\
		\hline
	\end{tabular}
\end{table*}

Table~\ref{tab:models} lists the
initial parameters and explosion diagnostics for all five models, following the definitions of
\citet{Kuroda2025}: $L_{\rm box}$ is the size of the computational domain (box); $J_0$ and $\Omega_0$ are the initial angular momentum and angular velocity;
$M_{\rm WD,bar}$ is the baryonic white-dwarf mass; $\beta_{\rm cb}$ is the same rotational-to-gravitational
energy ratio as $\beta_0$ (the same definitions are used as in Eqs.~(\ref{eq:Trot})-(\ref{eq:W})) but evaluated at core bounce (rather than in the initial hydrostatic
configuration), which is $\sim 5\beta_0$, reflecting the spin-up of the contracting core; $t_9$ is the post-bounce time at which
the maximum shock radius first reaches
$10^{9}\,{\rm cm}$; $t_{\rm fin}$ is the time at which each simulation is stopped, and $E_{\rm exp}$ is
the diagnostic explosion energy (the volume integral of the local specific energy over unbound matter
outside $100\,{\rm km}$, evaluated where the local binding energy is positive; Eq.~(4) of \citealt{Kuroda2025}), quoted at $t_9$ and at $t_{\rm fin}$.

We briefly outline the differences between the current models and those reported in \citet{Kuroda2025}.
First, we evolved all of our previous models for slightly longer times (e.g., $\sim60$\,ms for R2 and $\sim300$\,ms for R1).
This extension is needed to obtain more reliable asymptotic thermodynamic values after freeze-out from nuclear statistical equilibrium (NSE), which improves the nucleosynthesis calculation.
We also enlarged the Cartesian computational domain by a factor of 8 in each of the three directions ($x,y,z$), which results in a larger domain size of $L_{\rm box}=1.2\times10^5$\,km (used in R1, R6, and R6B11) from the previous value of $1.5\times10^4$\,km (used in models: R2 and R3), while the central finest resolution maintains $\Delta x\sim229$\,m throughout the models.
A brief description of the overall dynamics of the newly added model R6B11 can be found in Appendix~\ref{app:R6B11}.

Here, we note that in models R2 and R3, for which the computational domain extends to only $L_{\rm box}=1.5\times10^4$\,km, the shock front reaches the outer computational boundary at $t_{\rm pb}\approx380$\,ms, before the end of the simulations (see panel (c) of Fig.~\ref{fig:RshockEexp}). Consequently, some material leaving the computational domain after this time is not included in the measured ejecta mass. To assess the impact of this artificial loss on the following analysis, we estimate its magnitude relative to the total ejecta mass, $M_{\rm ej}\approx0.046$--$0.061\,M_\odot$, in these models. Using representative values of $\rho\approx10^3\,{\rm g\,cm^{-3}}$ and $v\approx0.1c$ at the outer boundary, $r=1.5\times10^9$\,cm, the mass-loss rate is estimated as $\dot{M}_{\rm out}\approx4\pi r^2\rho v\approx0.04\,{\rm M}_\odot\,{\rm s}^{-1}$.
The interval between the arrival of the shock at the outer boundary and the end of the simulation is approximately $38$--$65$\,ms. The corresponding artificially lost mass is therefore estimated to be $\Delta M_{\rm out}\approx(1.6\text{--}2.8)\times10^{-3}\,M_\odot$, which amounts to only a few per cent of the total ejecta mass.
We therefore neglect this artificial loss in the following discussion.

This approximation is further justified by the primary focus of this study. We are mainly interested in the hot ejecta launched from the vicinity of the central NS that have undergone NSE, rather than in the expelled outer layers of the progenitor WD, which constitute the main component crossing the outer boundary.
According to our previous study, the hot ejecta were still located at radii of only several thousand kilometres at the end of the simulation (see Fig.~6 of \citealt{Kuroda2025}), which cannot reach the outer boundary by the end of the simulation.
Thus, the vast majority of the ejecta relevant to the present nucleosynthesis analysis remains within the computational domain and is included in the post-processing calculations.

\subsection{Tracer particles and nucleosynthesis post-processing}
\label{sec:tracers}

For each model, Lagrangian tracer particles are placed in all numerical cells with positive gravitational binding energy, thus defined as unbound, at the final simulation time.
For each tracer particle, we assign a mass, which is the total mass contained in the corresponding numerical cell.
We then trace the trajectories of all particles backwards in time and record their thermodynamic evolution (density, temperature,
electron fraction, and position as functions of post-bounce time) \citep{FreiburghausRosswogThielemann2000,Seitenzahl2010,BovardRezzolla2017,Wanajo2014,Sieverding2023}.
Only the northern hemisphere ($z\geq0$, where $z$ is the rotation axis) is covered by tracers, consistent with the equatorial (reflection) symmetry imposed in the underlying simulations.
Whole-star totals (e.g.,\ the $^{56}$Ni-proxy masses of Table~\ref{tab:yields}) implicitly assume equatorial symmetry.

Tracer-particle post-processing is a well-established method for extracting detailed nucleosynthetic yields from multi-dimensional Eulerian simulations \citep{Travaglio2004}. Its accuracy can nevertheless depend on particle placement, mass weighting, temporal sampling, and the number of trajectories representing rare thermodynamic conditions \citep{Seitenzahl2010,Harris2017}. The backward-reconstruction strategy adopted here is particularly useful when inline tracers are unavailable because it preferentially reconstructs the evolution following the last NSE phase, although the yields of rare isotopes may remain sensitive to the snapshot cadence and tracer sampling \citep{Sieverding2023}. We therefore provide the sampling-completeness and solver diagnostics in Appendix~\ref{sec:appendix-diagnostics} and regard the sparsely sampled actinide-region component as non-robust.

We post-process each tracer trajectory with the nuclear reaction network WinNet \citep{Reichert2023}, whose tracer-plus-single-zone-network methodology was originally demonstrated by \citet{Winteler2012} for MR-SN jets, integrating to a final time of $1\,{\rm Gyr}$.
Our network comprises 6757 nuclei and is evolved with WinNet's implicit Euler solver.
Its standard framework is based on reaction rates in JINA REACLIB format \citep{Cyburt2010}, with theoretical Hauser--Feshbach rates and nuclear partition functions supplementing the experimentally constrained data \citep{Rauscher2000}.

A tracer is excluded by the density filter if any recorded density
is non-positive or if its density at the start of the trajectory is below $10^{3}\,{\rm g\,cm^{-3}}$, where the latter criterion is adopted to remove numerical artefacts arising from our atmosphere treatment.
Integrating every density-filtered tracer (several tens of thousands per model) with WinNet is
computationally prohibitive, so we select a uniformly spaced subset of the ordered list: every tenth
tracer for R1, R2, R3 and R6, and every fifth for R6B11, i.e.\ twice as many tracers for that model.
We denote this sampling stride by $k$ ($k=10$ and $k=5$, respectively).
The denser sampling for R6B11 is motivated by two properties that set it apart from the other four models: its longer post-bounce evolution ($t_{\rm fin}=764\,{\rm ms}$, roughly $1.2$--$1.8$ times longer than any other model; Table~\ref{tab:models}) and its substantially higher explosion energy ($E_{\rm exp,9}=6.01\times10^{50}\,{\rm erg}$, the highest in the sequence), both of which drive ejecta to more extreme, rarer conditions (Section~\ref{sec:massA}) that require finer tracer sampling to resolve statistically.
All mass-integrated quantities (e.g.,\ ejected mass versus mass number, Fig.~\ref{fig:massA}) are scaled by $k$ to recover an estimate of the full filtered population.

Because the sampling is systematic and uniform in the ordered tracer list, this recovers the correct \emph{shape} of the ejecta distributions; a further $0.03$--$13.7$~per cent of the \emph{sampled} tracers per model (largest for R6, see Appendix~\ref{sec:appendix-diagnostics}) fail to converge in WinNet and are excluded, which mildly and fairly uniformly (in composition) lowers the recovered ejecta mass for that model.
Table~\ref{tab:yields} summarises, for each model, the number of tracers actually computed, the recovered ejecta mass, and the fraction of ejecta reaching $T_{\max}\geq7\,{\rm GK}$ (``hot'' ejecta, i.e.\ matter that underwent NSE). For hot tracers, we additionally record the electron fraction $Y_e$ and the entropy per baryon $s$ at $7\,{\rm GK}$ ``freeze-out'', taken at the last downward crossing of $T=7\,{\rm GK}$ along each trajectory (i.e.,\ the last time the trajectory cools through this temperature, which matters for tracers that are transiently reheated), linearly interpolated between the two bracketing recorded time-steps; these freeze-out values, denoted $Y_e$ and $s$ throughout, underlie Figs~\ref{fig:ye}, \ref{fig:meridional_entropy} and~\ref{fig:equatorial}.
We emphasise that $T=7\,{\rm GK}$ is used here as an operational threshold for classifying the ejecta and recording their thermodynamic conditions, rather than as a universal, sharply defined NSE boundary.
As expanding matter cools, complete NSE generally gives way to quasi-statistical-equilibrium clusters before individual reactions freeze out, with the transition depending on density, electron fraction, and expansion timescale \citep{Hix1999,Meyer1998,Hix2006}.
The adopted value follows earlier WinNet post-processing studies of magnetorotational-supernova ejecta, which likewise assume NSE above $7\,\mathrm{GK}$ and switch to the full reaction network below this temperature \citep{Reichert2021,Reichert2023b}. A comparable criterion, $T_{\mathrm{NSE}}=0.6\,\mathrm{MeV}\simeq7\,\mathrm{GK}$, is also used in recent three-dimensional core-collapse-supernova nucleosynthesis calculations, where NSE additionally requires the strong-interaction timescale to be shorter than the density-evolution timescale \citep{WangBurrows2024}. Some studies instead adopt higher thresholds of approximately $9$--$10\,\mathrm{GK}$, which provide a more conservative initialisation in nearly fully photodisintegrated matter but do not define a universal NSE boundary. The hot/cold classification and freeze-out values quoted throughout this paper therefore refer specifically to our $7\,\mathrm{GK}$ criterion. Using the maximum temperature recorded along each trajectory, we have checked how the classification responds to a higher threshold. With $10\,\mathrm{GK}$, the hot fraction of Table~\ref{tab:yields} decreases by $4.7$--$6.7$ percentage points, to $78$--$90$~per cent. In every model, all of the beyond-iron-group material, $M(A\geq88)$, originates from tracers with $T_{\max}\geq10\,\mathrm{GK}$, so this part of our results does not depend on the choice between the two thresholds. The $^{56}$Ni-proxy mass is more sensitive: $13$--$36$~per cent of $M(A{=}56)$ comes from tracers with $7\leq T_{\max}<10\,\mathrm{GK}$. This test concerns the classification only; the network calculations, which switch between NSE and the full reaction network at $7\,\mathrm{GK}$, were not repeated with a higher switching temperature.

\begin{table}
	\centering
	\caption{Sampling and headline yield summary per model. $k$: tracer sampling stride
	(Section~\ref{sec:tracers}). $M_{\rm ej}$: recovered ejecta mass. Hot fraction: mass fraction with
	$T_{\max}\geq7\,{\rm GK}$.
	$M(A{=}56)$: ejected mass at mass number $A=56$, used here in place of the
	$^{56}$Ni mass (Section~\ref{sec:radioactive}). $M(A{\geq}88)$: recovered ejected mass beyond the iron
	group, i.e.\ in the first $r$-process-peak region and heavier.}
	\label{tab:yields}
	\footnotesize
	\setlength{\tabcolsep}{2.2pt}
	\begin{tabular}{lcccccc}
		\hline
		Model & $k$ & tracers & $M_{\rm ej}$ & hot & $M(A{=}56)$ & $M(A{\geq}88)$\\
		 & & computed & (M$_\odot$) & (per cent) & (M$_\odot$) & ($10^{-3}\,{\rm M}_\odot$)\\
		\hline
		R1 & 10 & 7204  & 0.0237 & 84.1 & 0.0068 & 0.097 \\
		R2 & 10 & 7633  & 0.0461 & 87.0 & 0.0095 & 0.391 \\
		R3 & 10 & 8145  & 0.0605 & 89.2 & 0.0100 & 0.769 \\
		R6 & 10 & 7919  & 0.0968 & 91.6 & 0.0140 & 1.728 \\
		R6B11 & 5  & 18979 & 0.1492 & 94.9 & 0.0142 & 8.029 \\
		\hline
	\end{tabular}
\end{table}

\subsection{Solar reference abundances}
\label{sec:solar}

Overproduction factors (Section~\ref{sec:overproduction}) are computed relative to the solar-system
abundance table of \citet{LoddersPalmeGail2009}, the same table shipped with WinNet and used by
\citet{Reichert2023b}, allowing a direct comparison of overproduction patterns between MR-SNe
and the AIC models studied here. This meteoritic/proto-solar compilation is not the only
standard in use; photospheric determinations \citep{Asplund2009,Asplund2021} give systematically lower
C, N and O abundances, which would shift the corresponding (already strongly underproduced, $A<20$)
overproduction factors slightly upwards without affecting our conclusions about the iron-group and
heavier nuclei.

\section{Results}
\label{sec:results}

\subsection{Ejected mass and composition}
\label{sec:massA}

\begin{figure}
	\includegraphics[width=\columnwidth]{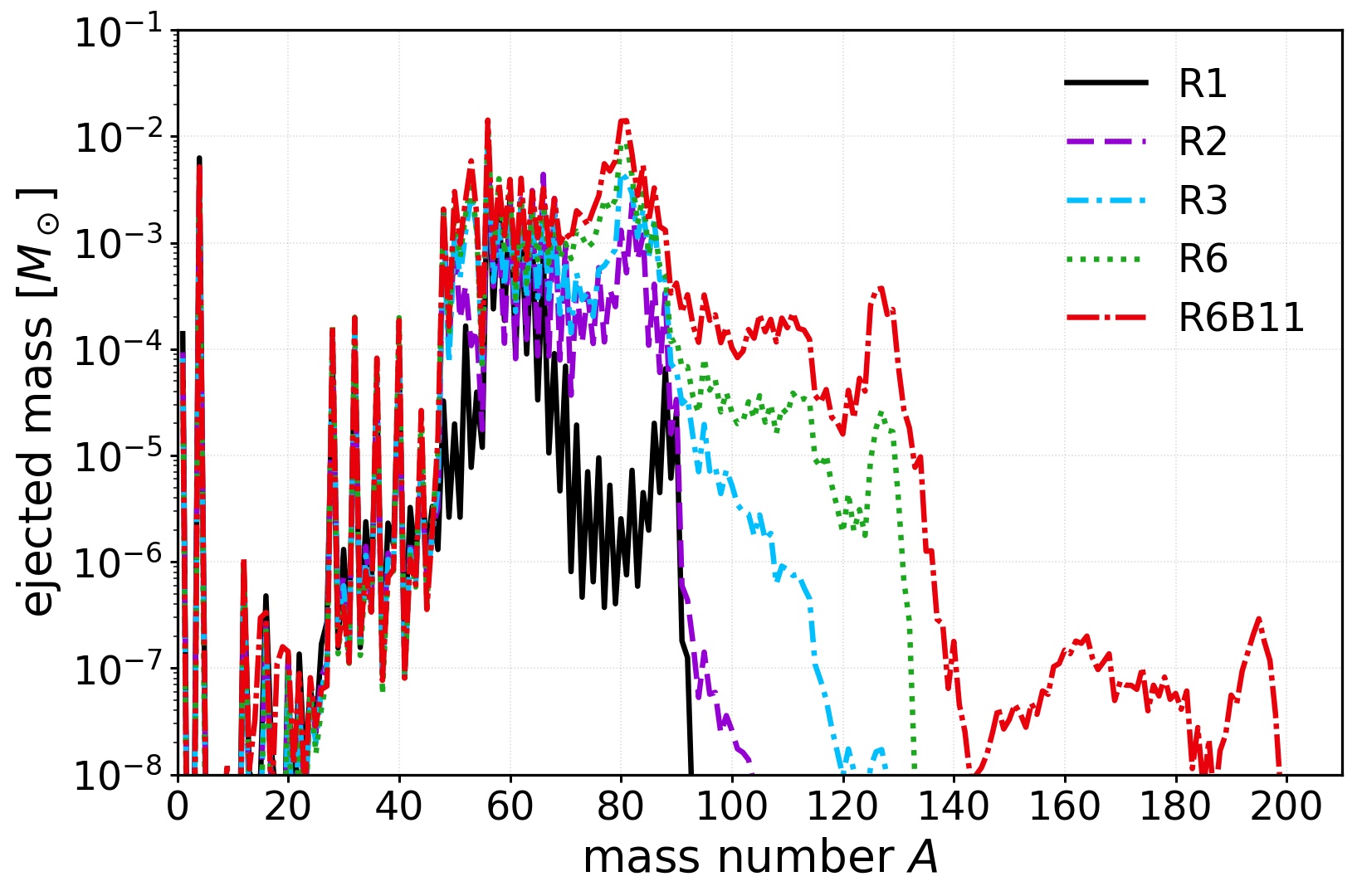}
\vspace{-5mm}
	\caption{Ejected mass versus mass number $A$ for all five models, scaled to the full
	density-filtered tracer population (Section~\ref{sec:tracers}).}
	\label{fig:massA}
\end{figure}

\begin{figure}
	\includegraphics[width=\columnwidth]{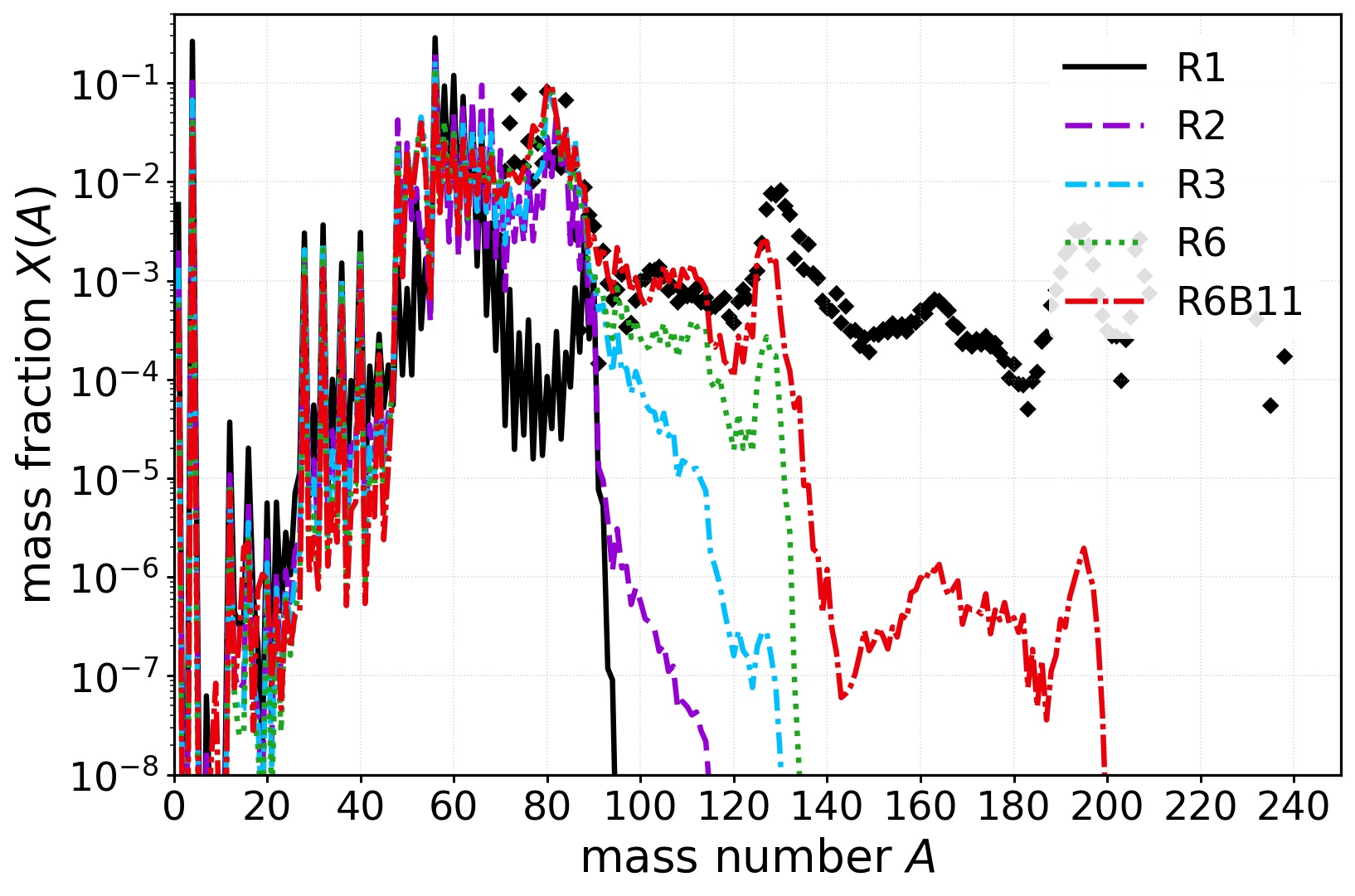}
\vspace{-5mm}
	\caption{Ejecta-averaged mass fraction $X(A)$ versus mass number, self-normalised (sampling-stride
	independent). Black diamonds show the solar $r$-process residual of \citet{Sneden2008}, as
	tabulated by \citet{Prantzos2020}, normalised at mass number $A=88$.}
	\label{fig:fracA}
\end{figure}

Figures~\ref{fig:massA} and~\ref{fig:fracA} show the ejected mass and ejecta-averaged mass fraction
versus mass number $A$ for all five models. The light/intermediate-mass $\alpha$-chain pattern (peaks at
$A=4,12,16,20,24,28,32,36,40,44$) is essentially identical in shape and relative amplitude across all
five models, indicating that this part of the yield is set by common outer-layer/incomplete-burning
conditions rather than by rotation rate or explosion energetics. Both contributions to this part of the yield are tied to the initial $Y_e=0.5$ of the WD: hot ejecta that are merely shock heated retain $Y_e\approx0.5$ (Section~\ref{sec:ye}), and the cold ejecta, which originate in the outer layers, are initialised with free nucleons in our network calculations (Section~\ref{sec:ye}). The amplitude of the light and intermediate-mass part of the pattern therefore depends in part on this initialisation; for example, cold tracers contribute $50$--$85$~per cent of the ejected mass at $A=44$ (Section~\ref{sec:spatial}). We therefore do not interpret the overproduction in this mass range quantitatively. All five models peak at the iron group
($A\approx56$), but this peak becomes progressively broader and less dominant with increasing rotation:
$X(A{=}56)\approx0.29$ for the slowest rotator, R1, about $1.4$ to $3$ times higher than for the
other four models ($X(A{=}56)\approx0.10$--$0.21$), with the difference redistributed into a broader
$A>56$ tail. R2, intermediate in rotation rate between R1 and R3, is correspondingly
intermediate in this peak fraction ($X(A{=}56)\approx0.21$), consistent with the monotonic trend.
The light $\alpha$-chain pattern and the iron-group peak are prominent features of the nucleosynthetic yields of ordinary massive-star supernovae, reflecting both the pre-supernova hydrostatic burning history and shock-induced explosive burning \citep[e.g.,][]{Rauscher2002}.
These yields provide a useful diagnostic of the relevant burning regimes.
By contrast, an AIC progenitor has neither a massive stellar mantle nor an extended envelope, so there is comparatively little contribution from the original component of progenitor WD to the eventual explosive nucleosynthesis outcomes. Newly synthesized material in the post explosion phase can therefore constitute a much larger fraction of the total AIC ejecta and exert a correspondingly stronger influence on the final relative abundance pattern. Any quantitative comparison between AIC and massive-star supernova yields must consequently account for this fundamental difference in progenitor structure and in the relative contributions of pre-existing and newly synthesized material to the ejecta.

The mass number up to which nucleosynthesis proceeds increases monotonically and substantially with
rotation, explosion energy, and magnetic fields: the bulk yield extends from $A\approx90$ (R1) through $A\approx115$--120
(R2) and $A\approx120$--135 (R3, R6) to $A\approx200$ (R6B11), and the maximum atomic number for which the yield is overproduced relative to solar abundances correspondingly rises from $Z\approx40$ (R1) through
$Z\approx50$ (R2) to $Z\approx50$--58 (R3, R6) to $Z\approx80$ (R6B11;
Section~\ref{sec:overproduction}). A
first-$r$-process-peak-like feature near $A\approx80$ ($N=50$)
appears in all but the slowest rotator, including R2; a second-peak-like feature near $A\approx130$
($N=82$) appears only in the two fastest models, an order of magnitude stronger in R6B11 than in
R6, and is absent (below the noise floor) in R2 just as in R1 and R3. R6B11
alone shows a marginal, noisy tail reaching the third $r$-process peak ($A\approx195$, $N=126$) and, at even
smaller
ejected masses ($\sim10^{-10}\,{\rm M}_\odot$, below the range plotted in Fig.~\ref{fig:massA}),
extending into the actinide region (mass numbers up to $A\approx244$). Because this tail is produced by
only a handful of tracers, we report it as a non-robust, trace-level signal rather than as evidence for
genuine third-peak or actinide nucleosynthesis; reaching even a marginal third-peak signal at all,
however, is itself notable, since it is achieved only by the single most extreme model in our sequence
(highest rotation rate \emph{and} magnetically boosted explosion energy; Section~\ref{sec:ye}).

The absolute mass of this beyond-iron-group material, $M(A\geq88)$, increases monotonically and steeply
with rotation and final explosion energy $E_{\rm exp,fin}$, from $9.7\times10^{-5}\,{\rm M}_\odot$ (R1) through
$3.9\times10^{-4}$, $7.7\times10^{-4}$ and $1.7\times10^{-3}\,{\rm M}_\odot$ (R2, R3,
R6) to $8.0\times10^{-3}\,{\rm M}_\odot$ (R6B11; Table~\ref{tab:yields}) -- an increase of nearly
two orders of magnitude across the sequence, faster than the factor of $\sim6$ growth in total ejecta
mass over the same range. Expressed as a fraction of the total ejecta, the beyond-iron-group mass
fraction itself rises monotonically from $0.4$~per cent (R1) to $5.4$~per cent (R6B11): faster and, in the case of R6B11, magnetically boosted models do not merely eject more mass overall; they eject a systematically larger \emph{share} of that mass as heavy, beyond-iron-group material.
This is the direct, absolute-mass counterpart to the heaviest-mass-number and overproduction-factor trends described above.

Comparing R6 and its magnetised counterpart R6B11, which share identical initial rotation
parameters (Table~\ref{tab:models}) but differ in the presence of a magnetically-driven outflow and in
the resulting explosion energy ($E_{\rm exp,fin}=4.27\times10^{50}$ versus $8.52\times10^{50}\,{\rm
erg}$), isolates magnetically-boosted explosion energy, and not rotation rate \emph{per se}, as a
comparably important control on how far nucleosynthesis proceeds beyond the iron group; at identical
rotation, the magnetised model ejects $4.6$ times more beyond-iron-group mass ($M(A\geq88)$) than its
purely hydrodynamic counterpart (Table~\ref{tab:yields}). This control is
corroborated by the spatial and $Y_e$ trends discussed in Sections~\ref{sec:ye} and~\ref{sec:spatial}.

\subsection{Overproduction factors}
\label{sec:overproduction}

\begin{figure*}
	\centering
	\includegraphics[width=0.95\textwidth]{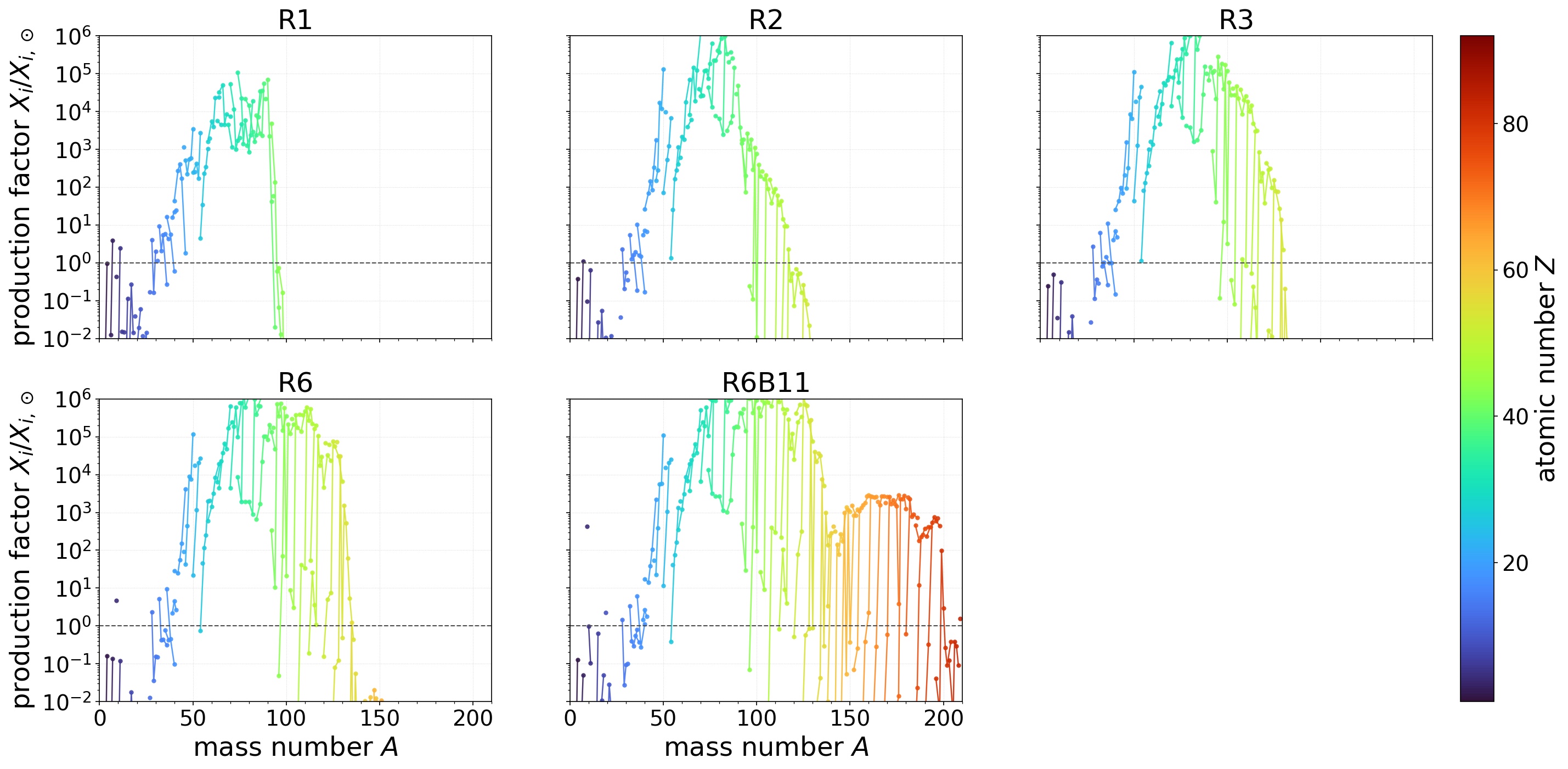}
	\caption{Production factor $X_i/X_{i,\odot}$ versus mass number, one panel per model, isotopes of
	the same element connected and coloured by atomic number $Z$.}
	\label{fig:overproduction}
\end{figure*}
Figure~\ref{fig:overproduction} shows the production factor relative to solar abundances,
$X_i/X_{i,\odot}$, versus mass number, one panel per model, computed for each isotope $i=(Z,A)$ as
$f_i=(m_i/M_{\rm ej})/(A\,Y_{i,\odot})$, where $m_i$ is the ejected mass of that isotope, summed over
tracers analogously to the mass-versus-$A$ distributions of Section~\ref{sec:massA}, and $Y_{i,\odot}$
is its solar mole fraction (Section~\ref{sec:solar}). Iron-group and light trans-iron nuclei
($A\approx50$--90) are overproduced by factors of $10^3$--$10^5$ in all five models, while the lightest
tracked species ($A\lesssim10$--20) are underproduced. The high-$A$/high-$Z$ edge of the overproduction
envelope scales with rotation and explosion energy in the same sequence identified in
Section~\ref{sec:massA}: from $Z\approx40$ (R1) through $Z\approx50$ (R2) to $Z\approx50$--58
(R3, R6) to a
distinct secondary plateau, at production factors of order $10^2$--$10^3$, extending to $Z\approx80$
(Hg) in R6B11 alone; a few actinide isotopes (Th, U, Pu) are additionally present in R6B11 at
near-solar production factors (a few times solar), corresponding to the trace tail noted in
Section~\ref{sec:massA}. This confirms, in production-factor form, that overproduction of heavy, high-$Z$
nuclei in this model set is unique to the fastest and most energetic model, with the remaining four
models, including R2, forming a smooth, monotonic progression below it.

For comparison, self-consistent two-dimensional neutrino-driven explosion models of electron-capture and low-mass core-collapse supernovae have found production factors of order $10^{2}$ for light trans-iron elements from $A\approx65$ (Zn) to $A\approx90$ (Zr), associated with moderately neutron-rich innermost ejecta, whereas the corresponding ejecta of more massive progenitors tend to be less neutron rich or proton rich \citep{Wanajo2018}. The extension of the overproduction plateau to substantially heavier nuclei in R6B11 therefore reflects its pronounced low-$Y_e$ component and magnetorotationally assisted mass ejection, rather than a generic feature of ordinary neutrino-driven core-collapse supernovae.

The positive slopes within individual isotopic chains, most prominently seen in R6B11, qualitatively resemble those of the neutron-rich MR-SN models of \citet{Reichert2023b}. Because the progenitor structures, ejecta masses, and explosion geometries differ substantially, we restrict this comparison to the qualitative abundance trend.
Many isotopes exhibit a positive slope in their abundance trends, indicating a neutron-rich environment.
This trend is particularly pronounced in the magnetised model R6B11.
It can be attributed to the magnetic fields, which enhance mass ejection and thereby reduce the dwell time of matter exposed to neutrino irradiation.
A similar positive-slope signature---indicative of a neutron-rich environment---is observed in other non-magnetised models, though the trend is less pronounced than in R6B11 and weakens with decreasing rotation rate.
Even in the non-magnetised sequence, relatively neutron-rich conditions arise due to centrifugally driven mass ejection.
In our earlier work \citep{Kuroda2025}, we demonstrated that non-axisymmetric instabilities efficiently eject material from the vicinity of the PNS core---where the electron fraction
$Y_e$ is low---while simultaneously reducing the contribution of neutrino heating prior to ejection, thereby suppressing the rise in
$Y_e$.
This rotational mechanism leads to neutron-rich mass ejection in non-magnetised AIC, which is consistent with the findings of \citet{Batziou2025}.
However, in our slowest-rotating model, R1, the abundances within an individual isotopic chain no longer rise systematically towards the more neutron-rich isotopes over the mass range $60 \lesssim A \lesssim 90$: the slopes flatten there and in places reverse sign.
This behaviour is primarily due to the relatively stronger role of neutrino heating in driving mass ejection in this model, leading to a proton-rich environment compared to faster-rotating cases.
The finite simulation time may also play a role. Long-term axisymmetric simulations show that rotating AIC models eject increasingly neutron-rich matter at later times, because an accretion torus feeds neutrino-driven bipolar outflows for several seconds \citep{Batziou2025}. R1 is evolved to $t_{\rm fin}=617$\,ms, longer than R2, R3 and R6 (Table~\ref{tab:models}), so the comparison within our sequence is not biased against R1; we cannot, however, exclude that R1 develops such neutron-rich outflows in a later phase that none of our simulations covers.

\subsection{Electron fraction and hot/cold ejecta}
\label{sec:ye}
\begin{figure}
	\includegraphics[width=\columnwidth]{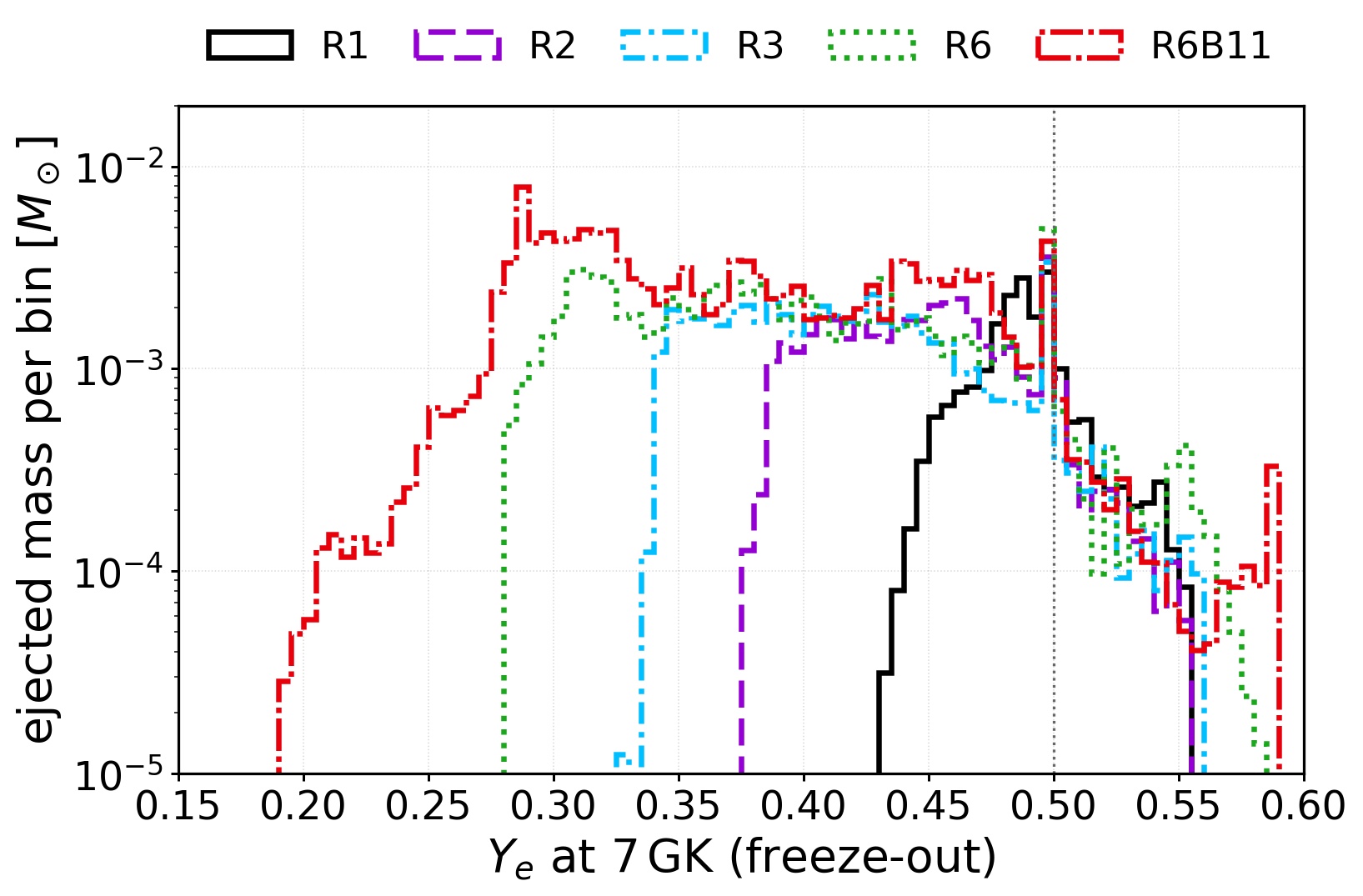}
	\caption{Mass-weighted electron fraction $Y_e$ at $7\,{\rm GK}$ freeze-out, hot ejecta only; tracers that never reach $7\,{\rm GK}$ have no defined freeze-out $Y_e$ and are excluded. Bin width $\Delta Y_e=0.005$.}
	\label{fig:ye}
\end{figure}
Figure~\ref{fig:ye} shows the mass-weighted electron-fraction ($Y_e$) distribution at $7\,{\rm GK}$ freeze-out for hot ejecta.
All five models show a common spike at $Y_e\approx0.50$ (the signature of initial $Y_e=0.5$ of progenitor WD, which does not experience strong neutrino absorption processes, but is simply shock heated).
As another noteworthy feature, the mass reaching more neutron-rich conditions increases monotonically with rotation rate: R1 remains confined to $Y_e\gtrsim0.42$; R2 extends only slightly further, with essentially no mass below $Y_e\approx0.38$; R3 reaches down to $Y_e\approx0.33$; R6 develops a substantial low-$Y_e$ shoulder peaking near $Y_e\approx0.31$ and extending to $Y_e\approx0.28$, with $4.4$~per cent of its hot ejecta below $Y_e=0.30$; and R6B11 is by far the most neutron-rich, peaking at $Y_e\approx0.29$ and extending in a tail to $Y_e\approx0.19$, with $19.8$~per cent of its hot ejecta below $Y_e=0.30$. In R6B11 the mass contained in this low-$Y_e$ peak ($0.26\leq Y_e<0.32$, $4.3\times10^{-2}\,{\rm M}_\odot$) exceeds that in the $Y_e=0.50$ spike ($0.48\leq Y_e<0.52$, $9.5\times10^{-3}\,{\rm M}_\odot$) by a factor of about $4.5$.
This $Y_e$ trend is the direct physical driver of the mass-number and production-factor
trends of Sections~\ref{sec:massA}--\ref{sec:overproduction}: lower freeze-out $Y_e$ enables progressively heavier, more neutron-rich nucleosynthesis products.
At fixed entropy and expansion time-scale, a decrease in $Y_e$ generally increases the neutron-to-seed ratio. The heaviest mass number ultimately attained, however, is jointly determined by $Y_e$, entropy, expansion time-scale, and neutrino interactions \citep{Arcones2011,Bliss2018}. The freeze-out $Y_e$ distribution should therefore be regarded as the principal, but not the sole, diagnostic of the increasingly heavy abundance patterns in our model sequence.

A comparison between R6B11 and R6 clearly indicates that the magnetic fields increase the neutron-rich ejecta mass.
A set of 2D axisymmetric magnetised AIC models reported by \citet{Cheong25_AIC} also presents a similar trend.
Although the numerical setup is completely different and a direct comparison with our results is not straightforward, one of their models using $B_0=10^{11}$\,G as the initial poloidal magnetic field strength retains ejecta with $Y_e$ in the range $0.25$--$0.5$, whereas their most strongly magnetised model ($B_0=10^{12}$\,G) reaches even lower values, $Y_e\lesssim0.25$ and in places as low as $Y_e\approx0.1$. It is nevertheless notable that our R6B11, which adopts the same initial field strength $B_0=10^{11}\,$G, peaks at $Y_e\approx0.29$ and extends down to $Y_e\approx0.19$, consistent with the neutron-rich range they report for that model.

Because R6 and R6B11 share the same initial rotation, this pair shows directly how the magnetic field acts on the nucleosynthesis. In R6B11 the magnetically driven outflow doubles the explosion energy and increases the ejecta mass by 54~per cent (Tables~\ref{tab:models} and~\ref{tab:yields}; Appendix~\ref{app:R6B11}). The main nucleosynthetic consequence is a shift of the hot ejecta towards low $Y_e$: the fraction of hot ejecta below $Y_e=0.30$ rises from $4.4$ to $19.8$~per cent, and the lowest freeze-out $Y_e$ falls from $\approx0.28$ to $\approx0.19$ (Fig.~\ref{fig:ye}). We attribute this to the faster magnetically assisted ejection, which shortens the time the matter is exposed to neutrino irradiation before freeze-out (Section~\ref{sec:overproduction}).
Although we have not performed the same analysis, \cite{Reichert2024} reported that low-$Y_e$ materials can be ejected via strong magnetic fields within the first $\sim100$\,ms after bounce before they become more proton rich through neutrino absorption.
The heavy-element yields follow this shift: $M(A\geq88)$ increases by a factor of $4.6$, the second-peak feature near $A\approx130$ by an order of magnitude, and only R6B11 shows a tail towards the third peak (Section~\ref{sec:massA}). The $^{56}$Ni-proxy mass, by contrast, is almost unchanged ($0.0140$ versus $0.0142\,{\rm M}_\odot$), consistent with its production in near-symmetric matter (mass-weighted $Y_e\approx0.47$--$0.48$ of the hot ejecta producing it in the two models) rather than in the neutron-rich component. Spatially, the additional neutron-rich matter of R6B11 is not ejected along the magnetically driven polar outflow but, like the heavy elements, at low and intermediate latitudes (Section~\ref{sec:spatial}).

\begin{figure}
	\includegraphics[width=\columnwidth]{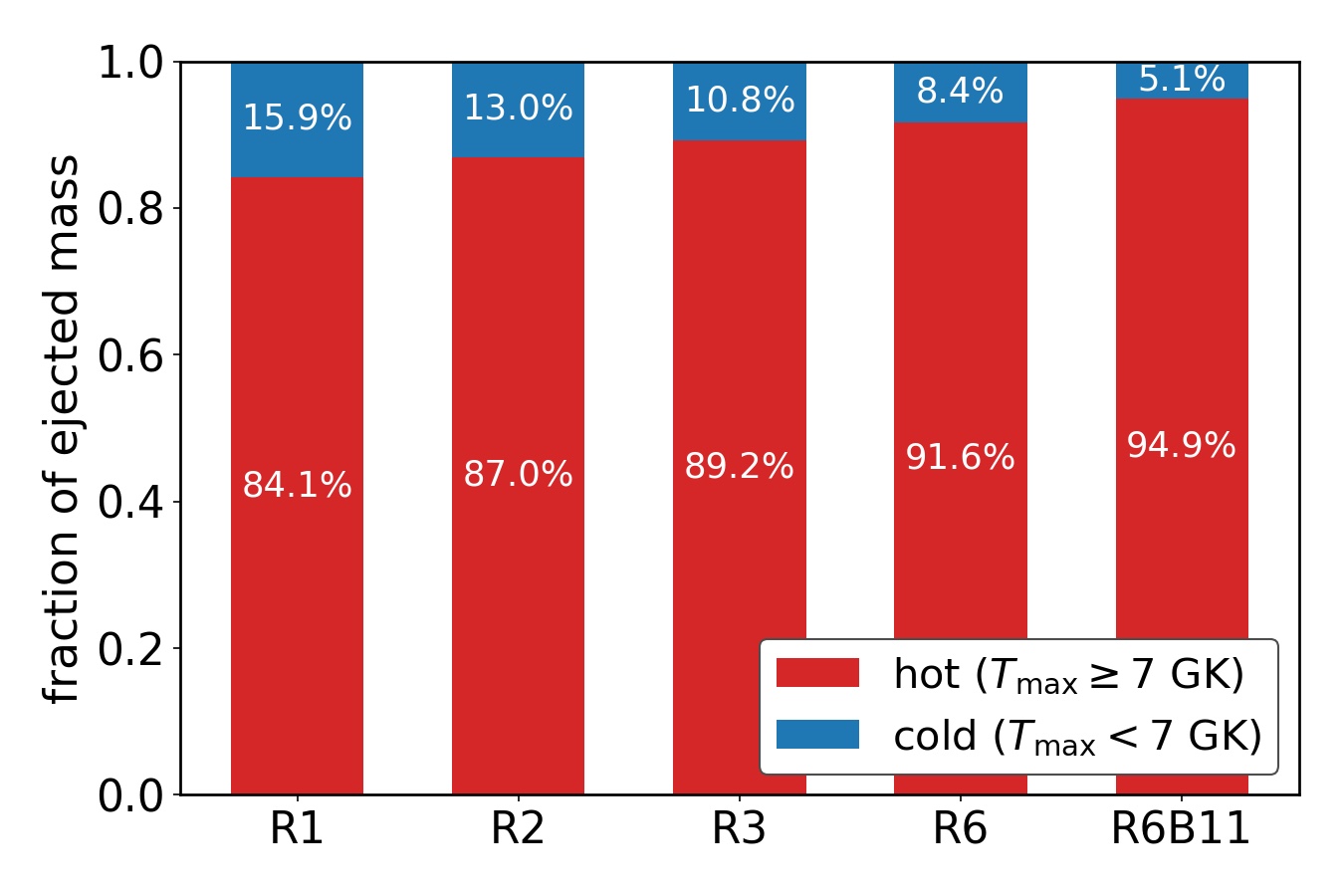}
	\caption{Hot ($T_{\max}\geq7\,{\rm GK}$) versus cold ejected-mass fraction per model.}
	\label{fig:hotcold}
\end{figure}
Figure~\ref{fig:hotcold} shows the corresponding hot ($T_{\max}\geq7\,{\rm GK}$) versus cold
ejected-mass fraction. The hot fraction increases monotonically with rotation rate, from $84.1$~per cent
(R1) through $87.0$, $89.2$ and $91.6$~per cent (R2, R3, R6) to $94.9$~per cent
(R6B11;
Table~\ref{tab:yields}), with a corresponding decrease in the cold ejecta fraction.
It is noteworthy that the model R1 is our second-longest simulation (see Table~\ref{tab:models}), implying that the larger fraction of cold ejecta does not stem from a shorter simulation time, but instead is a clear signature of its weaker explosion with less heated matter.
This trend is robust across the rotation sequence and is further strengthened in the magnetised model.

The cold components are ejected without undergoing NSE.
Consequently, the final composition obtained from the WinNet calculation depends on the initial WD surface composition, since these components originate primarily from the outer layers of the progenitor WD.
In the current study, although our EOS table gives an initial composition of $^{28}$Si in the vicinity of the surface, as it is simply the component used to construct our tabulated SN EOS in the low-density and low-temperature regimes, we adopt a simple default setup provided by the original WinNet: free nucleons.
Ideally, nucleosynthesis calculations for these cold ejecta should be initialised with a more physically motivated composition, which we leave for future work.
Although this approach simplifies the treatment of the initial composition, we do not consider it to compromise the beyond-iron-group results of this paper.
This simplification is justified because the cold component is a minor part of the ejecta, amounting to $15.9$, $13.0$, $10.8$, $8.4$ and $5.1$~per cent of the ejected mass for R1, R2, R3, R6 and R6B11, i.e.\ becoming least important precisely in the models with the largest heavy-element yields. More importantly, the cold component contributes no mass at all to $M(A\geq88)$ in any of the five models, so the beyond-iron-group results of Sections~\ref{sec:massA} to~\ref{sec:spatial} are independent of the assumed initial composition of these tracers.
This robustness does not, however, extend to the lighter $^{56}$Ni-proxy mass $M(A{=}56)$
(Section~\ref{sec:radioactive}), to which the cold component contributes non-negligibly. A dedicated
sensitivity test on a mass-representative sample of cold R6B11 tracers, comparing the adopted
free-nucleon initial composition against a physically motivated pre-collapse ONeMg white-dwarf
composition (e.g.\ $X(^{16}{\rm O})\approx0.55$, $X(^{20}{\rm Ne})\approx0.28$,
$X(^{23}{\rm Na})\approx0.06$, $X(^{24}{\rm Mg})\approx0.05$; \citealt{GilPons2001}), shows that this
choice increases the mass-integrated $M(A{=}56)$ by $\approx30$~per cent and $M(A{=}60)$ by roughly a
factor of five for that subset. Since the cold ejecta contribute of order $34$--$46$~per cent of the total
$M(A{=}56)$ budget across our models, this could shift the reported $^{56}$Ni-proxy totals of
Table~\ref{tab:yields} by about $10$~per cent. We therefore flag the initial composition of cold
ejecta as a genuine methodological uncertainty affecting $M(A{=}56)$, $M(A{=}44)$ and $M(A{=}60)$, and
defer a dedicated production rerun with the corrected seed composition to future work.

\subsection{Radioactive isotopes}
\label{sec:radioactive}

In this section, we focus on the following directly observable radioactive isotopes $^{56}$Ni, $^{44}$Ti, $^{60}$Fe and $^{26}$Al: $^{56}$Ni
(via its daughter $^{56}$Co) powers supernova light curves and sets the classification of brightness of transient events (normal supernovae/hypernovae/AICs); $^{44}$Ti is detected directly in young supernova remnants through its decay
gamma-ray lines \citep{Grebenev2012,Siegert2015}; and $^{60}$Fe and $^{26}$Al are mapped as diffuse
Galactic gamma-ray line emission tracing ongoing nucleosynthesis \citep{Diehl2006,Wang2007,Wang2020}.
Because \texttt{finab.dat} is written at the network's final integration time ($1\,{\rm Gyr}$), the short-lived radioactive isotopes of interest ($^{56}$Ni, $t_{1/2}=6.1\,{\rm d}$; $^{44}$Ti, $t_{1/2}=60\,{\rm yr}$) have fully decayed.
We therefore use the (conserved) mass at the corresponding mass number as a proxy for the synthesised radioactive yield ($M(A{=}56)\to{}^{56}$Ni, $M(A{=}44)\to{}^{44}$Ti, $M(A{=}60)\to{}^{60}$Fe, $M(A{=}26)\to{}^{26}$Al); this reproduces the correct \emph{spatial distribution} and relative trends across models but not an absolute yield directly comparable to literature values quoted at one-tenth of a half-life, as commonly tabulated \citep[e.g.][their table~3]{Reichert2023b}.

Table~\ref{tab:yields} lists the $^{56}$Ni-proxy mass for each model. The yield increases only
weakly and monotonically with rotation, from $\approx0.007\,{\rm M}_\odot$ (R1) through
$\approx0.0095\,{\rm M}_\odot$ (R2) to
$\approx0.014\,{\rm M}_\odot$ (R6B11), about a factor of two across the full sequence, despite
the ejecta mass increasing by a factor of $\sim6$ and the hot fraction rising from 84 to 95~per cent
(Table~\ref{tab:yields}). All five models fall far below both the typical core-collapse-supernova
$^{56}$Ni yield of $\sim0.07\,{\rm M}_\odot$
\citep{Seitenzahl2014,Kristoffer2009,Anderson2019}, for which observed values span $\approx0.005$--$0.18\,{\rm M}_\odot$ in normal Type~II supernovae with a mean of $0.037\,{\rm M}_\odot$ \citep{Rodriguez2021} and the mean over all core-collapse types is $0.055\,{\rm M}_\odot$ \citep{Rodriguez2023}, and the hypernova threshold of
$\sim0.1$--0.3$\,{\rm M}_\odot$ \citep{Nomoto2006,NomotoKobayashiTominaga2013}, by factors of five to ten and seven to forty-three, respectively (comparing our full $0.007$--$0.014\,{\rm M}_\odot$ range against each benchmark), implying that the $^{56}$Ni-powered component of the AIC light curve is much fainter than in ordinary core-collapse supernovae at every rotation rate in our sequence. This constrains the radioactively powered emission only; a central engine can still make the early light curve bright (Section~\ref{sec:discussion}).
Modern 3D simulations of neutrino-driven core-collapse supernovae show that the $^{56}$Ni yield correlates with the explosion energy, with $\approx0.002$--$0.26\,{\rm M}_\odot$ over explosion energies of $\approx(0.1$--$2.5)\times10^{51}$\,erg \citep{Burrows2024}. Our least energetic model, R1 ($E_{\rm exp,fin}=1.2\times10^{50}$\,erg), lies within the range found there for comparably weak explosions, whereas our more energetic models fall increasingly below that relation; R6B11 ($8.5\times10^{50}$\,erg) yields about seven to nine times less $^{56}$Ni than their models of similar explosion energy.
We note that jet collimation itself can, in principle, enhance $^{56}$Ni production
\citep{MaedaNomoto2003}; the weak $^{56}$Ni response to rotation found here suggests that this effect is subdominant in our AIC models compared with the much smaller absolute ejecta mass available to burn to the iron group.
This weak dependence may also partly stem from the non-axisymmetric rotational instability, as it works to uncollimate the bipolar outflows, thereby tending to reduce the fractional contribution of rotation to the $^{56}$Ni production.
In our previous study \citep{Kuroda2025}, we have crudely estimated the $^{56}$Ni mass and obtained $M_{\rm Ni}=2.6(2.0)\times10^{-3}$\,$M_\odot$ for R6(R1), which were used for the discussion of $^{56}$Ni powered light curve.
These previous rough estimates increase by factors of $\sim3$--$5.5$ after a more sophisticated nucleosynthesis calculation as well as longer simulation times. This comparison refers to the $^{56}$Ni mass only, the sole isotope estimated in \citet{Kuroda2025}.

\subsection{Spatial distribution: entropy and heavy-element correlation}
\label{sec:spatial}

\begin{figure*}
	\centering
	\includegraphics[width=\textwidth]{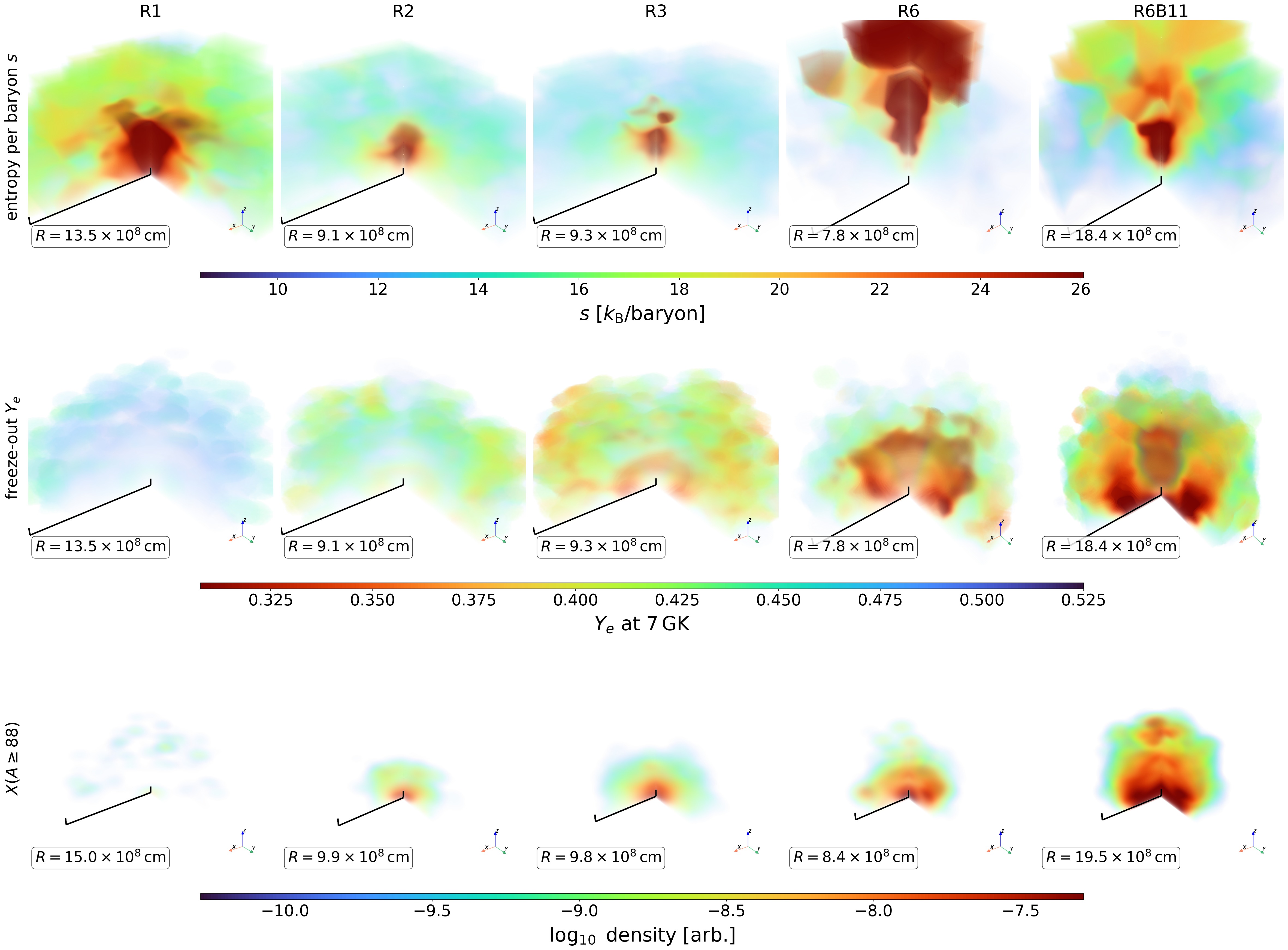}
	\caption{Three-dimensional volume rendering (octant wedge cut) of entropy per baryon $s$ (top row, in
$k_{\rm B}$\,baryon$^{-1}$), mass-weighted freeze-out electron fraction $Y_e$ of the hot ejecta (middle row) and $X(A\geq88)$ density (bottom row, $\log_{10}$ density in arbitrary units)
for all five models at the final simulation time. Each row is shown on a single colour scale common to all five models
(one colour bar per row), so colours are directly comparable across models. The labelled arrow
in each panel marks the radius from the rotation axis to the outer extent of the plotted material
along the equatorial direction, in units of $10^{8}\,$cm (entropy-per-baryon and $Y_e$ rows: hot-ejecta extent; $X(A\geq88)$ row: extent of the $A\geq88$ material). In the $Y_e$ row, low $Y_e$ is rendered opaque and $Y_e\approx0.5$ transparent, and cells without appreciable hot-ejecta mass are left transparent.}
	\label{fig:meridional_entropy}
\end{figure*}
Figure~\ref{fig:meridional_entropy} shows three-dimensional (octant-wedge-cut) renderings of entropy per baryon, freeze-out $Y_e$ and $X(A\geq88)$, for all
five models, constructed by binning the sparse tracer set of each model onto a $96^3$ Cartesian grid (individually sized to that model's own 99.9th-percentile tracer extent, with a  margin of 8 per cent in the equatorial plane and 30 per cent along the rotation axis) and
applying a Gaussian smoothing kernel ($\sigma=2.4$ grid cells) to obtain a continuous volume-rendered
field from the discrete tracers. The box, and hence the radius indicated in each panel, therefore differs between the models. It follows the extent of each model's ejecta at the end of its simulation, which is set by the different simulation times ($t_{\rm fin}=418$--$764$\,ms; Table~\ref{tab:models}) and ejecta velocities (Fig.~\ref{fig:RshockEexp}). We use model-specific boxes so that the $96^3$ grid resolves the structure of every model equally well; a common box would confine the most compact models to a small fraction of the grid. A high-entropy, collimated column along the rotation axis is present in all
five models and becomes progressively taller and more sharply collimated with increasing rotation rate,
consistent with rotational/centrifugal collimation of the outflow; the \emph{peak} entropy per baryon
itself, however, does not scale monotonically with rotation (the maxima of the smoothed field
span $\approx29$--$37\,k_{\rm B}$/baryon with no clear trend; note that the colour scale shared by all
five panels saturates at $26\,k_{\rm B}$/baryon), indicating that rotation primarily controls the
outflow's geometric collimation rather than its peak value of the entropy per baryon.

Contrary to the picture established for jet-driven $r$-process outflows in MR-SNe
\citep{Winteler2012,Nishimura2015}, the heaviest nucleosynthetic yields do not track this high-entropy
polar column: in every model the heaviest ejecta ($X(A\geq88)$, $X(A>100)$, $X(A>150)$ -- three
progressively more restrictive thresholds spanning the first $r$-process peak region and probing towards
the, largely unreached, second and third peaks of Section~\ref{sec:massA}) instead reside
in equatorial-to-mid-latitude, moderate-entropy lobes, while the narrow polar column itself is a
\emph{local minimum} in heavy-element mass fraction (Fig.~\ref{fig:meridional_entropy}). This holds without exception for nuclei with $A>100$: within $10^\circ$ of the rotation axis, their mass fraction is below 5~per cent of the ejecta-averaged value in every model that produces them. The less restrictive threshold $A\geq88$ has one exception, R6, in which this material is concentrated at intermediate latitudes but is not depleted near the axis. Its polar column, which carries $1.5$~per cent of the ejected mass, consists of high-entropy ($s\approx30\,k_{\rm B}$\,baryon$^{-1}$), slightly proton-rich ($Y_e\approx0.55$) matter whose $A\geq88$ mass fraction is about three times the ejecta average, although it contains essentially no nuclei with $A>100$. Because Fig.~\ref{fig:meridional_entropy} shows the mass density rather than the mass fraction, this low-density material is barely visible there. This
anti-correlation, rather than the positive correlation naively expected from the neutron-to-seed-ratio
argument that high entropy favours a heavier $r$-process at fixed $Y_e$
\citep{WoosleyHoffman1992,Meyer1992,HoffmanWoosleyQian1997}, indicates that in these AIC models the
modest heavy-element yield originates from the moderate-entropy equatorial outflow rather than from the
high-entropy jet itself, more reminiscent of the low-entropy, low-$Y_e$ route to a heavy $r$-process found
in neutron-star-merger ejecta \citep{FreiburghausRosswogThielemann1999, Wanajo2014} than of the high-entropy jet route.
This concentration of the heaviest nucleosynthetic products at equatorial-to-mid-latitudes is the outcome of the non-axisymmetric rotational instability that efficiently ejects low-$Y_e$ components from the PNS core surface \citep{Kuroda2025}.
Although the entropy is not so high there (see the upper row in Fig.~\ref{fig:meridional_entropy}), the relatively low-$Y_e$ conditions enhance the rapid neutron capture process to reach the 2nd $r$-process peak.
The freeze-out $Y_e$ (middle row of Fig.~\ref{fig:meridional_entropy}) shows this directly: in the faster rotators, the most neutron-rich hot ejecta sit at low and intermediate latitudes near the base of the outflow, where the heavy-element material resides, whereas the matter near the rotation axis has $Y_e$ close to or slightly above $0.5$.
$X(A>150)$ is absent in R1, R2 and R3, negligible in R6 ($\sim10^{-12}\,{\rm M}_\odot$ in total), and appreciable only in R6B11 ($\approx4\times10^{-6}\,{\rm M}_\odot$), but even there this material originates from the equatorial/low-latitude base of the
outflow, not from the polar column.

\begin{figure*}
	\centering
	\includegraphics[width=\textwidth]{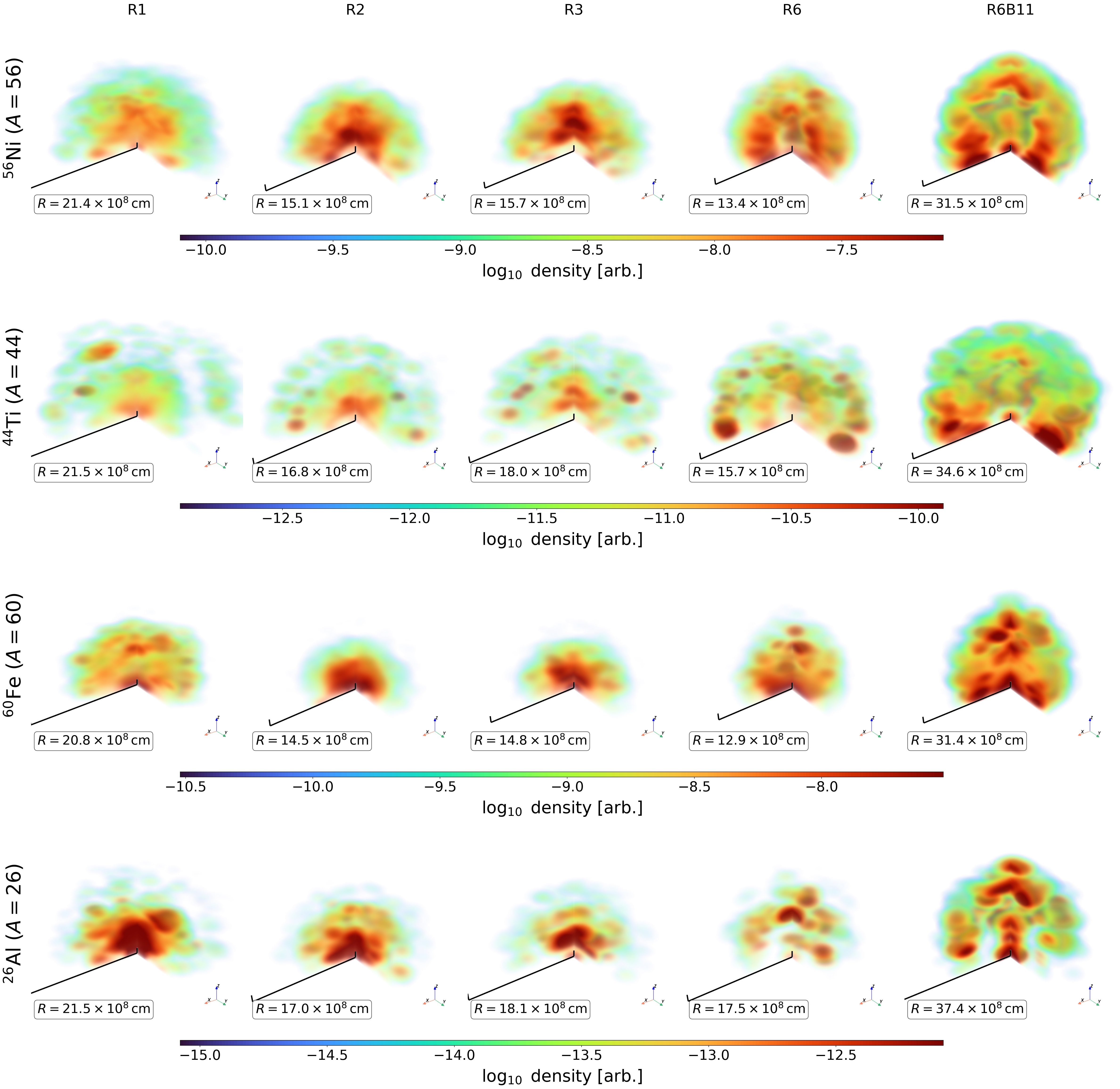}
	\caption{Three-dimensional volume rendering (octant wedge cut) of the four radioactive-isotope
proxy densities (Section~\ref{sec:radioactive}) for all five models; colour scale is
$\log_{10}$ density in arbitrary units, shown on a single scale common to all five models per
isotope (one colour bar per row), so colours are directly comparable across models. The labelled arrow in each panel marks the radius from the rotation axis to the outer extent of that isotope's material along the equatorial direction, in units of $10^{8}\,$cm.}
	\label{fig:meridional_isotopes}
\end{figure*}
Figure~\ref{fig:meridional_isotopes} is the same as Fig.~\ref{fig:meridional_entropy} but for the four isotope-proxy densities.
We find that $^{56}$Ni and $^{44}$Ti trace
similarly broad envelopes extending to the largest radii and heights, including off-axis, jet-adjacent
lobes, whereas $^{60}$Fe and $^{26}$Al are more confined to a narrower, near-axis, low-altitude region,
suggesting these two isotope groups sample a more restricted, likely more neutron-rich, inner subset of
the ejecta than the broader $\alpha$-rich/iron-group material traced by $^{56}$Ni and $^{44}$Ti.

Strongly asymmetric and clumpy distributions of radioactive material are not unique to magnetorotational explosions. The asymmetric $^{44}$Ti distribution observed in Cassiopeia~A and 3D neutrino-driven core-collapse-supernova simulations both demonstrate that large-scale hydrodynamic asymmetries can strongly affect the relative distributions of $^{44}$Ti and $^{56}$Ni \citep{Grefenstette2014,Wongwathanarat2017}. More recent self-consistent 3D nucleosynthesis calculations further show that non-monotonic, long-time thermodynamic histories can substantially enhance $^{44}$Ti production relative to one-dimensional estimates \citep{Sieverding2023b}. The distinctive result of the present AIC models is therefore not asymmetry by itself, but the systematic rotation-dependent transition towards equatorially concentrated $^{56}$Ni- and $^{44}$Ti-proxy material.

The overall ejecta morphology transitions
systematically with rotation, from a smooth, filled, quasi-spherical cloud with no central cavity
(R1), through an intermediate, increasingly core-concentrated cloud (R2), a dense core with a
hint of a bipolar gap (R3), to a clear hollow bipolar/axial
cavity surrounded by a dense equatorial torus (R6), to the most pronounced double-lobe,
hourglass-like structure with a wide evacuated central channel (R6B11); i.e.\ faster rotation (and,
for R6B11, additional magnetic driving) produces a progressively more jet-like, hollowed-out
nucleosynthesis geometry even though, as noted above, the heaviest material itself avoids the evacuated polar channel.
We compare the ejecta morphology of our radioactive-isotope proxies with that reported by \citet{Reichert2023b}.
A major difference is the inhomogeneous distribution of $^{56}$Ni and $^{44}$Ti, concentrated at equatorial-to-mid-latitudes.
\citet{Reichert2023b} present a clear bipolar signature in all isotope distributions, basically for all models except one model ``W'' (see their Fig.~15).
Their model ``W'' employs weak initial magnetic fields that tend to suppress the formation of canonical magnetorotationally driven bipolar outflows, thereby resulting in launching shock blobs in all directions including along the equator.
Such shock expansions are analogous to our models, which are driven primarily by centrifugal forces.
Therefore we consider that the concentration of some heavy nuclei in the equatorial plane is a robust feature of rotating AICs and SNe that are non-magnetised or only weakly or moderately magnetised. Among AIC simulations, a strong latitude dependence of the electron fraction was already found in the 2.5D models of \citet{Dessart2006}; we compare our results with other AIC simulations in more detail in Section~\ref{sec:discussion}.

\begin{figure*}
	\includegraphics[width=\textwidth]{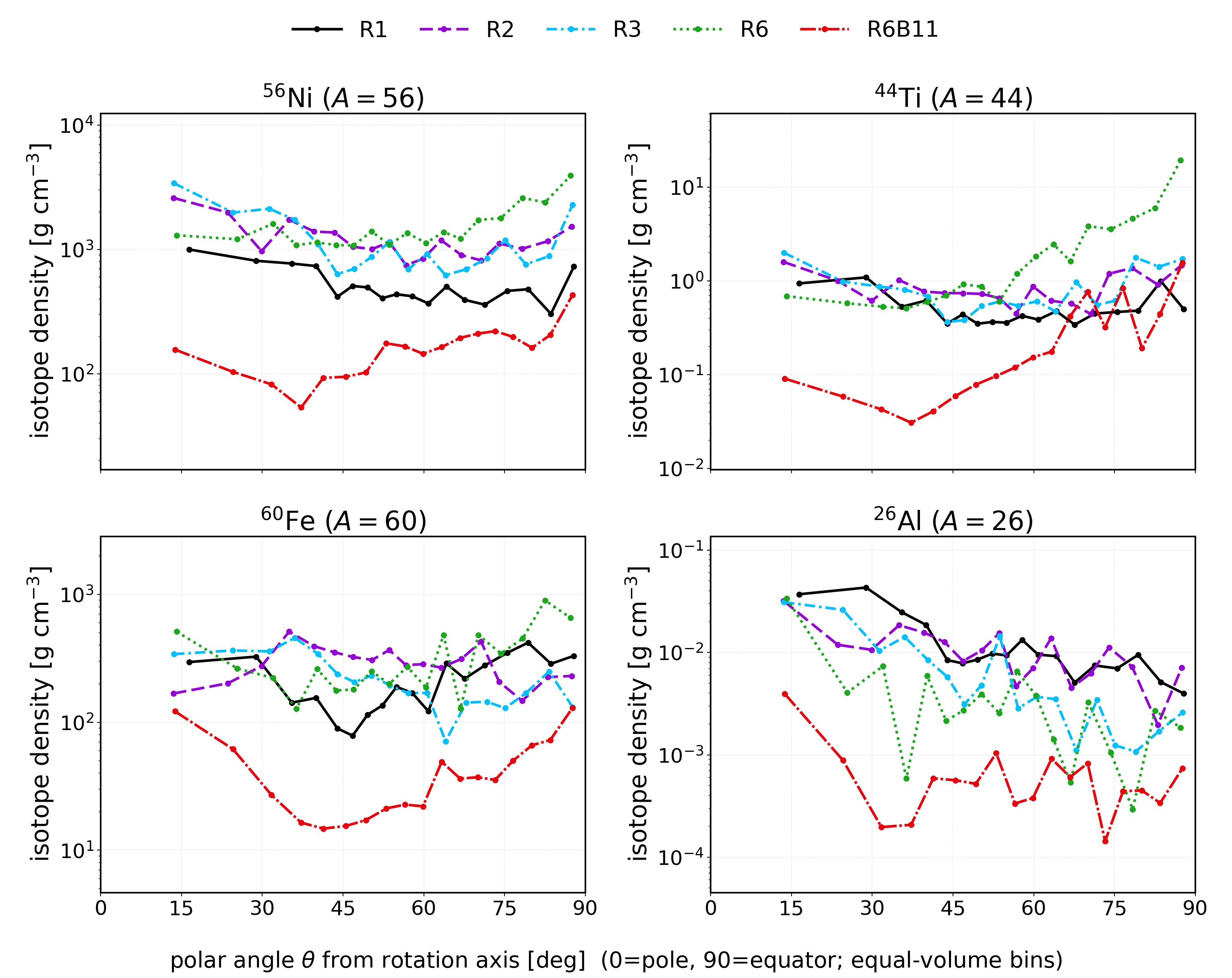}
	\caption{Local isotope density $\rho_{\rm iso}$ (mass-weighted, using an individual volume $V_i=m_i/\rho_i$ per
	tracer) versus polar angle $\theta$ from the rotation axis ($\theta=0^{\circ}$: pole;
	$\theta=90^{\circ}$: equator), in 18 equal-total-volume bins per model, all five models
	overlaid.}
	\label{fig:equatorial}
\end{figure*}

Figure~\ref{fig:equatorial} quantifies the angular distributions of the unstable isotopes shown in Fig.~\ref{fig:meridional_isotopes}.
For this purpose, we first divide the current simulation domain (i.e., northern hemisphere with $z>0$) between the rotation axis and the equatorial plane into 18 equal volume bins.
The volume-averaged mass density of each isotope in a bin is then estimated from the tracer particles in the corresponding bin as $\rho_{\rm iso}=\sum_i m_iX_i/\sum_i V_i$, where $V_i=m_i/\rho_i$, and $m_i$, $X_i$, and $\rho_i$ are the tracer mass, isotope mass fraction, and final hydrodynamic density, respectively.
This equal-volume binning reduces the risk of empty or poorly sampled bins that can arise from an arbitrary division, such as uniform angular binning, particularly near the rotation axis.
In addition, as the simulation time differs for all models, the final tracer-particle densities also differ, i.e., the longer the simulation time is (e.g., R6B11), the lower the ejecta density becomes.
Therefore, the slopes of the individual profiles, rather than their absolute values, are physically meaningful.

The resulting density profiles reveal a genuine, rotation-dependent \emph{reversal} for $^{56}$Ni and $^{44}$Ti.
In the three slower-rotating models (R1, R2, R3), the density near the pole is comparable to that at the equator or slightly higher by $1.1$--$1.9$ times.
In the two fastest, most energetic models (R6,
R6B11), this reverses: the equator is $3$ times denser than the pole in $^{56}$Ni, and $17$--$28$ times denser in $^{44}$Ti. This reversal is a direct, quantitative counterpart to the morphological transition
described above: the slower models have not yet developed a hollowed-out polar channel, so their densest
material simply sits close to the rotation axis, whereas R6 and R6B11 have opened a genuine low-density
polar funnel whose evacuated core is bypassed by the density-weighted $^{56}$Ni and $^{44}$Ti
distribution, which instead peaks in the surrounding equatorial torus. $^{26}$Al breaks from this pattern
entirely: it remains denser at the pole than at the equator in \emph{all five} models, by a factor of
$4.5$--$18$, with no reversal at high rotation, decoupling its spatial distribution from whatever process
opens the polar channel for the other proxies. $^{60}$Fe shows no consistent trend with rotation
(pole-to-equator density ratios of $0.7$--$2.6$, without a clear ordering by model), and we do not read a
robust angular signal into its profile.

The angular profiles observed in the plot show that the
pole/equator dichotomy is not a fixed geometric property but a consequence of the rotation-driven
opening of the polar channel: at low rotation there is no evacuated funnel for the densest $^{56}$Ni and
$^{44}$Ti to avoid. That $^{26}$Al does not follow this transition, remaining pole-concentrated at every
rotation rate, indicates that whatever opens the channel to the other proxies does not displace the
$^{26}$Al-forming material. Given its low and noisy yield we regard this as a genuine but
lower-confidence result, meriting confirmation with denser tracer sampling; for $^{60}$Fe we read no
robust angular signal at our present resolution.

These angular trends can be traced back to the freeze-out conditions, using the mass-weighted $Y_e$, entropy per baryon and polar angle of the hot tracers that produce each proxy. The $^{26}$Al-proxy material forms preferentially in high-entropy, near-symmetric matter ($s\approx18$--$23\,k_{\rm B}$\,baryon$^{-1}$ and $Y_e\approx0.45$--$0.51$, compared with mean entropies of $\approx12$--$19\,k_{\rm B}$\,baryon$^{-1}$ for all hot ejecta), and this matter lies closer to the rotation axis (mean polar angle $\approx29^\circ$--$44^\circ$, compared with $\approx55^\circ$--$58^\circ$ for all hot ejecta), consistent with the pole concentration of $^{26}$Al in every model. The hot part of the $^{56}$Ni- and $^{44}$Ti-proxy material originates from near-symmetric matter ($Y_e\approx0.47$--$0.50$ across the five models) with entropies comparable to or below the mean, and a large fraction of both proxies comes from cold ejecta; their distribution therefore follows the bulk of the ejected mass, which concentrates in the equatorial torus once the polar channel opens. The $^{60}$Fe proxy is produced almost entirely in hot ejecta (more than 94~per cent) under moderately neutron-rich conditions ($Y_e\approx0.42$--$0.48$), and its producers have a mean polar angle ($\approx48^\circ$--$60^\circ$) close to that of the hot ejecta as a whole, consistent with the absence of a clear angular trend. The angular structure of the freeze-out conditions themselves is visible in the top and middle rows of Fig.~\ref{fig:meridional_entropy}: in the two fastest models, the mass-weighted $Y_e$ of the hot ejecta decreases from $\approx0.48$--$0.50$ in the bin closest to the pole to $\approx0.34$--$0.36$ at intermediate latitudes, while the entropy decreases from $\approx21$--$22$ to $\approx10\,k_{\rm B}$\,baryon$^{-1}$ towards the equator.

Because the cold ejecta carry an uncertain initial composition (Section~\ref{sec:ye}), we repeated
this analysis using hot tracers only. The reversal for $^{56}$Ni and $^{44}$Ti persists in both fast
models; for $^{56}$Ni it in fact strengthens, the equator-to-pole density ratio rising from $\approx3$
for the full sample to $\approx9$ (R6) and $\approx13$ (R6B11). $^{26}$Al likewise remains
pole-concentrated in all five models. The numerical factors are, however, sensitive to this cut: for
$^{44}$Ti, to which the cold component contributes $50$--$85$~per cent of the mass, the equator-to-pole
ratio of the two fast models falls to $\approx6$--$11$. The \emph{direction} of the angular trends is
therefore robust against the treatment of the cold ejecta, whereas the individual factors should be read
as order-of-magnitude estimates.

\section{Discussion}
\label{sec:discussion}

We now place these results in the context of other AIC simulations, of the jet $r$-process mechanism, of the chemical
enrichment budget, and of the electromagnetic signatures expected from AIC.

Other recent AIC simulations differ from ours in dimensionality, magnetisation and simulated time, but they report the ejecta properties that set the nucleosynthesis and can be compared directly. The 2.5D non-magnetised models of \citet{Dessart2006} produced weak explosions, with $3$--$4\times10^{49}$\,erg and a few $10^{-3}\,{\rm M}_\odot$ of ejecta more than $600$\,ms after bounce. Our non-magnetised 3D models eject considerably more mass ($0.024$--$0.097\,{\rm M}_\odot$) with explosion energies of $(1.2$--$4.3)\times10^{50}$\,erg. Differences in progenitor, rotation and neutrino transport preclude a one-to-one comparison; in our models, the non-axisymmetric instabilities, which redistribute angular momentum, additionally assist the mass ejection \citep{Kuroda2025}. The fully general-relativistic 3D, non-magnetised simulations of \citet{LongoMicchi2023} likewise found AIC to be electromagnetically faint, about two orders of magnitude fainter than Type~Ia supernovae, in line with our low $^{56}$Ni-proxy masses. The long-term 2D simulations of \citet{Batziou2025} show that rotating models eject increasingly neutron-rich matter for several seconds through torus-fed bipolar outflows, a phase not covered by our simulations (see below).

With strong magnetic fields, the ejecta become considerably more neutron-rich. The 2D magnetohydrodynamic models of \citet{Dessart2007} ejected $\sim0.1\,{\rm M}_\odot$ with $Y_e=0.1$--$0.2$ in explosions of a few $10^{51}$\,erg. In the 2D general-relativistic neutrino-magnetohydrodynamic simulations of \citet{Cheong25_AIC} and \citet{Pitik2026}, the models with the strongest initial field, $10^{12}$\,G, eject strongly neutron-rich matter; \citet{Pitik2026} obtain $\approx0.2\,{\rm M}_\odot$ with $\langle Y_e\rangle\sim0.24$ and $r$-process nucleosynthesis up to and beyond the third peak. Our magnetised model R6B11, with $B_0=10^{11}$\,G, ejects $0.15\,{\rm M}_\odot$ with a freeze-out $Y_e$ distribution that peaks at $\approx0.29$ and reaches the second but not robustly the third $r$-process peak, consistent with the $Y_e$ range that \citet{Cheong25_AIC} report for the same initial field strength (Section~\ref{sec:ye}). Taken together, these results indicate that the strength of the initial magnetic field is a key parameter for the heavy-element yield of AIC. \citet{Pitik2026} find the bulk of their neutron-rich material ($Y_e\lesssim0.25$) in what they call the mid-polar region, whereas the neutron-rich, heavy-element-bearing ejecta of our models reside at low and intermediate latitudes. A comparison of the angular structure across field strengths and between axisymmetric and 3D dynamics requires 3D simulations at stronger fields or with self-consistent field amplification, such as those of \citet{Combi2025}, in which a magnetorotational dynamo amplifies an initially weak field, launches a jet and drives a neutron-rich disc wind.

That no model synthesises a robust third $r$-process peak is a markedly weaker outcome than the
full $r$-process, including actinides, obtained in the most favourable MR-SN jet
models \citep{Winteler2012,Nishimura2015}. It is, however, consistent with the caution established from
genuine 3D (rather than axisymmetric) magnetorotational simulations, in which a 3D kink instability can
disrupt the jet and strongly suppress $A>130$ ejection unless the pre-collapse magnetic field is
unrealistically strong \citep{Mosta2018}: even our most favourable, magnetically driven model (R6B11)
appears to sit well within the ``weak'' regime of that mechanism.
A useful neighbouring comparison is provided by electron-capture supernovae of super-AGB
stars, which share the collapse of an ONeMg core but differ from AIC in their progenitor envelope and
angular-momentum structure. Detailed calculations of these events generally favour trans-iron and first-peak
production rather than a robust $r$-process synthesis~\citep{Wanajo2009,Wanajo2011}.

State-of-the-art MR-SN simulations further show that the
nucleosynthetic outcome depends not only on the field strength but also on its topology and orientation
\citep{Reichert2024}. This sensitivity, together with the dependence of jet formation on rotation and
magnetic-field amplification \citep{ObergaulingerAloy2017}, motivates extending the present AIC
sequence beyond a single magnetised initial configuration.

The spatial anti-correlation between the heaviest products and the high-entropy polar column
(Section~\ref{sec:spatial}) suggests that in AIC, unlike in MR-SNe, the polar
channel is primarily a route for launching low-density, high-entropy but nucleosynthetically
unremarkable material, while the modest $r$-process-like signature we do find originates in denser,
moderate-entropy equatorial ejecta. The R6/R6B11 pair isolates magnetic driving from rotation, the
larger explosion energy being one of its principal dynamical consequences. Both findings should be
tested against independent AIC simulations and against stronger seed fields than the single magnetised
case considered here.

The monotonic shift of the ejecta towards heavier mass numbers at lower freeze-out $Y_e$ is consistent
with parametrised reaction-network studies, in which $Y_e$, entropy, and expansion time-scale jointly
determine whether the ejecta remain near the iron group, form a weak $r$-process, or reach the
lanthanides \citep{LippunerRoberts2015}. Neutrino absorption can shift this outcome by changing $Y_e$
before weak freeze-out \citep{ArconesThielemann2013}, making the competition between rapid ejection and
neutrino exposure particularly important for the angular abundance structure found here.

The long-term AIC study of \citet{Batziou2025} reported that rotating AIC/merger-induced-collapse models develop an early proton-rich, late neutron-rich $Y_e$ evolution (the opposite trend to non-rotating models) sustained for several seconds by a torus-fed neutrino-driven wind, and suggested this late-time neutron-rich phase as a candidate $r$-process channel.
Our short-term ($\lesssim1\,{\rm s}$), fully three-dimensional tracer-network calculation shows that even the fastest, most energetic rotator in our sequence only marginally and non-robustly approaches the third $r$-process peak within this time window.
We, however, emphasise that our short-term models do not exclude the torus-fed neutron-rich outflow occurring at a much later phase identified by \citet{Batziou2025}.
Whether it would push our yields further towards a genuine $r$-process is a natural extension of this work, but is beyond the duration of the simulations post-processed here.

Given the heavy-element masses reported above, it is worthwhile to assess the cumulative
contribution of AIC ejecta to Galactic chemical evolution in light of the estimated event rate.
To date, no transient has yet been unambiguously identified as an AIC event, and the AIC occurrence rate therefore remains poorly constrained observationally \citep{Wang2020b}.

Our slowest-rotating model, R1, ejects $\sim1.0\times10^{-4}\,{\rm M}_{\odot}$ of nuclei with $A\geq88$, whereas the corresponding yield of R6B11 is $\sim8.0\times10^{-3}\,{\rm M}_{\odot}$, approximately 80 times larger (see Table~\ref{tab:yields}).
Using the total Galactic AIC rates quoted in Section~\ref{sec:intro}, namely $(1.7$--$9.8)\times10^{-3}\,{\rm yr}^{-1}$ when all channels are included and $(0.6$--$4.7)\times10^{-3}\,{\rm yr}^{-1}$ when the uncertain double-CO-WD channel is excluded \citep{Liu2020,Wang2020b}, and assuming that the $A\geq88$ yields of the various progenitor channels are bracketed by those of R1 and R6B11, the corresponding mass-injection rate over the combined rate and yield limits is
\begin{equation}
\dot{M}_{A\geq88}
\simeq
\left(
6.0\times10^{-8}
\mbox{--}
7.9\times10^{-5}
\right)
{\rm M}_{\odot}\,{\rm yr}^{-1}.
\label{eq:aic_mass_injection}
\end{equation}
For comparison, \citet{Rosswog2024} estimated that reproducing the present galactic chemical enrichment requires a time-averaged production rate
of approximately
$1.6\times10^{-6}\,{\rm M}_{\odot}\,{\rm yr}^{-1}$ when the full
$r$-process abundance distribution is included, and approximately
$2.5\times10^{-7}\,{\rm M}_{\odot}\,{\rm yr}^{-1}$ for the heavy
$r$-process component with $A>130$.
The former is broadly consistent with the rate--yield product of
$\sim10^{-6}\,{\rm M}_{\odot}\,{\rm yr}^{-1}$ implied by the merger rate and ejecta mass adopted by
\citet{VandeVoort2015} for neutron-star-merger enrichment.
These values should be regarded as galactic chemical evolution requirements rather than as the production rate of any particular astrophysical source; the contributions from neutron-star mergers, rare massive-star explosions, AIC events, and any other relevant sites must collectively satisfy this budget.

This comparison should be regarded only as an order-of-magnitude mass estimate.
The quantity $M_{A\geq88}$ is a simple sum over all nuclei with $A\geq88$, irrespective of their detailed isotopic distribution or production channel, and therefore does not represent an ejecta mass distributed according to the solar $r$-process abundance pattern.
In addition, the population distribution of AIC progenitor rotation rates and magnetic-field strengths, the nucleosynthetic yields of different progenitor channels, and the time dependence of the event rate are all uncertain.
Nevertheless, the limiting cases explored here indicate that AIC events could make a non-negligible contribution to the galactic chemical evolution of nuclei with $A\geq88$.

The modest $M(A=56)$ proxy masses obtained here suggest---if the $A=56$ material was initially dominated by $^{56}$Ni---that the radioactively powered component of the AIC transient should be substantially fainter than an ordinary core-collapse supernova.
This does not, however, preclude a bright early optical transient.
The rapidly rotating, magnetised AIC models such as R6oB of \citet{Kuroda2025}, the octant-symmetric counterpart of R6B11, can form a magnetar-strength proto-neutron star, whose spin-down energy, after diffusion through the low-mass ejecta, may produce an initial FBOT-like peak on a time-scale of approximately $2$--$3$ d as estimated by our previous study \citep{Kuroda2025}; related magnetar-powered AIC scenarios have also been proposed for AT~2018cow \citep{Lyutikov19,Yu2019}.
This estimation can be naturally applied to the current R6B11 model that is free from octant-symmetry constraint.
Radioactive heating associated with this $A=56$ material---provided that it was initially dominated by $^{56}$Ni---operates concurrently but may become observationally important only after the much brighter magnetar-powered component has declined.
The revised $M(A=56)$ proxy masses obtained in this work may therefore power a fainter late-time component and, if its effective diffusion time is sufficiently longer than that of the engine-powered emission, may give rise to a secondary maximum or shoulder in the light curve.
Although distinct from the dynamical ejecta studied here, neutrino-processed AIC disc winds have also been predicted to synthesise up to a few times $10^{-2}\,{\rm M}_{\odot}$ of $^{56}$Ni and power a radioactive transient peaking on a time-scale of approximately one day \citep{Metzger2009}; subsequent radiation-transfer calculations confirmed that such transients should be faint and rapidly evolving \citep{Darbha2010}.
Whether the two components appear as distinct peaks or as a single peak followed by a radioactive tail will depend on the magnetar spin-down and thermalisation efficiencies, gamma-ray diffusion, ejecta opacity, and the three-dimensional distribution of $^{56}$Ni.
Quantitative predictions based on the angular ejecta structure and nucleosynthetic yields obtained here are deferred to future multi-dimensional radiation-transfer calculations.

Other promising signatures include radio transients produced by the interaction of the ejecta with circumstellar material \citep{Moriya2016}.
Moreover, AIC models with substantially stronger magnetic fields than those considered here have been predicted to power kilonova-like emission \citep{Pitik2026}. Our present 3D model set, which contains only one moderately magnetised case, does not probe this strongly magnetised regime.

\section{Conclusions}
\label{sec:conclusions}

We have post-processed tracer particles from five 3D general-relativistic AIC simulations%
, four taken from \citet{Kuroda2025} and one newly developed magnetised model, spanning a sequence of initial rotation rates, with the WinNet nuclear reaction
network, following the analysis approach of \citet{Reichert2023b}. Our main conclusions are:

\begin{enumerate}
\item Ejecta mass ($0.024$--$0.149\,{\rm M}_\odot$), the hot-ejecta fraction ($84$--95~per cent), the mass number to which the bulk abundance distribution extends ($A\approx90$ to $A\approx200$), and the absolute ejected
mass beyond the iron group ($M(A\geq88)\approx1\times10^{-4}$ to $8\times10^{-3}\,{\rm M}_\odot$, nearly
two orders of magnitude) all increase
monotonically with initial rotation rate across our model sequence; the intermediate-rotation model
R2 falls cleanly between R1 and R3 on every one of these metrics. Faster-rotating models and, independently, the magnetised model R6B11 therefore do not merely eject more mass; they also eject a larger \emph{fraction} of that mass in the form of nuclei with $A\geq88$.
\item The corresponding freeze-out electron fraction extends to progressively more neutron-rich values
with increasing rotation, driving the transition from iron-group-dominated ejecta (slow rotation) to
weak first- and second-$r$-process-peak-like features (fast rotation); a robust third $r$-process peak is
not obtained in any of our five models.
\item The $^{56}$Ni mass, quantified by $M(A{=}56)$, increases only weakly with rotation
($0.007$--$0.014\,{\rm M}_\odot$) and is lower, in an order-of-magnitude comparison, than representative $^{56}$Ni yields inferred for
core-collapse supernovae in all models.
The radioactively powered component is therefore expected to be substantially fainter than an ordinary core-collapse supernova, but may emerge as a secondary maximum or late-time shoulder after an initial, much brighter magnetar-powered FBOT-like peak. Whether these components are temporally resolved will depend on their respective heating and effective diffusion time-scales.
\item Comparing R6 with its magnetised counterpart R6B11, which share identical initial
rotation but differ in explosion energy by a factor of two, shows that magnetically boosted explosion
energy is a control on the heaviest mass number attained that is comparable in importance to the rotation rate itself.
\item Contrary to the jet $r$-process picture established for MR-SNe, the heaviest
nucleosynthetic products in our AIC models are spatially anti-correlated with the high-entropy polar
outflow, instead residing in equatorial-to-mid-latitude, moderate-entropy ejecta.
However, we do not intend to claim that this difference arises solely from the distinct dynamics of AICs and canonical MR-SNe. Rather, we suggest that the onset of non-axisymmetric rotational instabilities and their impacts on the overall dynamics may depend on the progenitor structure, thereby leading to more neutron-rich ejecta towards the equator in the present AIC models.
\item The pole-versus-equator density distribution of individual isotope proxies is itself
rotation-dependent rather than fixed: $^{56}$Ni and $^{44}$Ti are pole-concentrated in the three
slower models but become equator-concentrated, by up to a factor of $\sim28$ for $^{44}$Ti, in the two
fastest models, tracking the transition from a filled ejecta cloud to a hollowed-out polar channel;
$^{26}$Al remains pole-concentrated at every rotation rate, decoupled from this transition.
\end{enumerate}

Taken together, these results suggest that AIC, at least for the non-magnetised and moderately magnetised, short-duration simulations considered here, is
not a robust site for a strong (third-peak/actinide) $r$-process, in contrast to the most favourable
MR-SN jet models, but does produce a modest, rotation- and energy-dependent
weak $r$-process signature whose spatial origin differs qualitatively from the polar-jet mechanism usually
invoked for this class of events.
We tentatively suggest that a solar-like $r$-process abundance pattern would be reproduced by AIC only in
the presence of an extremely strong magnetic field in the progenitor WD, well above the value adopted
here, although it is not clear how such strong field configurations would be established in nature.

\section*{Acknowledgements}
The authors declare no conflicts of interest.

LT and TK are grateful to Kyohei Kawaguchi for fruitful discussions on the electromagnetic radiation from magnetars.
TK also acknowledges Almudena Arcones, Yu An, and Jan Kuske for their valuable advice on using WinNet and for their warm hospitality at TU Darmstadt.
Numerical computations were carried out on Sakura and Raven at Max Planck Computing and Data Facility.
This work was in part supported by Grant-in-Aid for Scientific Research (No. 23H04900) of Japanese MEXT/JSPS.

\section*{Data Availability}
The data underlying this article will be shared on reasonable request to the corresponding author.

\bibliographystyle{mnras}
\bibliography{refs}

\appendix
\section{Newly developed model: R6B11}
\label{app:R6B11}
\begin{figure*}
\begin{center}
\includegraphics[angle=-90.,width=\textwidth]{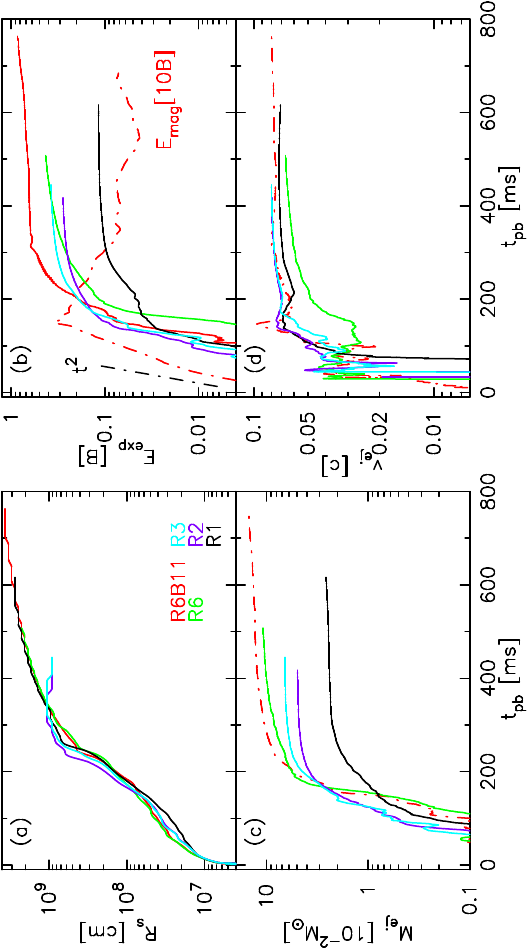}
\caption{Evolution of the maximum shock radius $r_{\rm shock}$ (top-left), diagnostic explosion energy $E_{\rm exp}$ (top-right), ejecta mass $M_\mathrm{ej}$ (bottom-left), and bulk velocity of ejecta $v_\mathrm{ej}$ (bottom-right) for all models. Note that the box size of R2 and R3 is $L_{\rm box}=1.5\times10^9$\,cm. Because of this and also of our post-process shock surface finding analysis, the shock front for these two models seems to stall at $\sim10^9$\,cm. However, the shock does pass through the outer boundary in the simulation.
In panel (b), we plot a supplemental line showing the magnetic energy of the system $E_{\rm mag}$ only for the magnetized model R6B11 (red dash-dotted line) together with a $t^2$-reference line (black dash-dotted line).
\label{fig:RshockEexp}}
\end{center}
\end{figure*}

Fig.~\ref{fig:RshockEexp} plots the time evolution of the maximum shock radius $R_{\rm s}$ in panel (a), diagnostic explosion energy $E_{\rm exp}$ and magnetic energy $E_{\rm mag}$ (only for our magnetized model R6B11) (b), ejecta mass $M_{\rm ej}$ (c), and ejecta velocity $v_{\rm ej}$ (d), where we employ the same definitions as in \citet{Kuroda2025}.
Since R1, R2, R3, and R6 have already been thoroughly discussed in our former study \citep{Kuroda2025}, here, we focus mainly on the dynamics of the newly developed model R6B11 and its differences from its counterpart non-magnetised model R6.
The evolution of maximum shock radius between these two models is essentially the same.
After bounce, the initial non-zero poloidal magnetic fields are wound up and amplified into a magnetically-driven bipolar outflow (diagnosed by the plasma beta $b^2/2P$, measuring the ratio of magnetic ($b^2/2$) to matter pressure ($P$), exceeding order-unity values along the polar funnel), rather than the purely hydrodynamic/neutrino-driven R6.

In the initial explosion phase ($t_{\rm pb}\lesssim250$\,ms), during which the shock still exists inside the progenitor WD (its typical radius is $\sim2\times10^8$\,cm, \citealt{Kuroda2025}) or just immediately after the shock breakout, we observe only a marginal difference in the shock position.
These deviations observed in the shock trajectories among different models become even more minor, once the shock reaches a sufficiently large distance from the WD (e.g., $r \gtrsim10^9$\,cm).
Although the shock evolution differs little between R6B11 and R6, we find a larger impact of magnetic fields on the explosion energy $E_{\rm exp}$ and ejecta mass $M_{\rm ej}$.
$E_{\rm exp}$ for R6B11 exhibits a value nearly twice as large and mass ejection also increases by $\sim50$~per cent.
The higher explosion energy and larger ejecta mass of R6B11 relative to R6 (see also Table~\ref{tab:models}) are naturally attributed to additional magnetic driving on top of the shared rotational energy budget.
This picture can be supported from the evolution of total magnetic energy $E_{\rm mag}$ (see red dash-dotted line in panel (b)). After bounce $E_{\rm mag}$ grows nearly as $t^{2}$ (i.e. the field strength grows nearly linearly) primarily via rotational winding as can be understood from a comparison with the $t^{2}$ line (black dash-dotted line). Although this initial growth rate is slightly suppressed after $t_{\rm pb}\sim50$\,ms, $E_{\rm mag}$ continues increasing and reaches $\sim3$\,B at $t_{\rm pb}\sim150$\,ms.
After reaching its peak value, $E_{\rm mag}$ significantly decreases, meanwhile $E_{\rm exp}$ for R6B11 rapidly increases.

Among the set of non-magnetised models (R1, R2, R3 and R6), the averaged ejecta velocity ($v_{\rm ej}\sim \sqrt{2E_{\rm exp}/M_{\rm ej}}$) of R6 (green line in panel (d)) shows the slowest speed.
This trend is primarily due to its larger ejecta mass.
With the presence of magnetic fields, however, the explosion energy substantially increases by a factor of $\sim2$, which is more efficient than the ejecta mass increase in this magnetised model R6B11.
Consequently, R6B11 has a higher ejecta velocity (red line in panel (d)) compared to R6, though the terminal ejecta velocity is still comparable to those of the other models (R1, R2, and R3).

Regarding the comparison with our previous octant symmetry model R6oB (though not shown in the present paper), the overall dynamics of the two models are broadly similar, with the sole distinction lying in the emergence of lower-order non-axisymmetric instability modes in R6B11.
As in other models (R1, R2, R3 and R6), the dominant non-axisymmetric flow patterns are one-armed ($m$=1) spiral waves.
These non-axisymmetric flow structures primarily modify the explosion dynamics in the equatorial plane, predominantly through angular momentum redistribution.
This transport mechanism spins up the outer envelope of the central proto-neutron star (PNS) \citep{Ott05,Kuroda2025}, thereby facilitating shock expansion in R6B11.
Meanwhile, the shock expansion along the polar direction (i.e., parallel to the rotation axis) remains largely unaffected, showing minimal divergence between the two models.

\section{Sampling completeness and solver diagnostics}
\label{sec:appendix-diagnostics}

For R6, the model with the largest solver attrition, the full tracer processing chain is: 98,179
total tracers $\rightarrow$ 91,742 pass the density filter of Section~\ref{sec:tracers} (6,437 removed
for negative or sub-threshold initial density, which can appear in our post-process passive particle transport, particularly at the sharp density profile in the vicinity of the WD surface) $\rightarrow$ 9,176 sampled at stride $k=10$ for WinNet
integration $\rightarrow$ 7,919 successfully computed, a 13.7~per cent failure rate among sampled
tracers (Table~\ref{tab:yields}); the corresponding failure rate is $\lesssim0.55$~per cent for the other
four models (0.03, 0.05, 0.17 and 0.53~per cent for R1, R2, R3 and R6B11
respectively). The dominant WinNet failure mode ($\approx85$~per cent of failures) is error code
W430008, ``hydro time-step too small'', with minor contributions from wall-clock timeouts, negative
expansion velocity, and mass non-conservation errors. Failures are clustered in contiguous bursts along
the ordered particle-ID sequence (i.e.\ along contiguous trajectories). Within R6 itself, the
failure rate is uniform across its computational domain, rather than concentrated in one particular region,
indicating a systematic limitation of the stock WinNet adaptive-time-step solver on rapidly varying,
shock-adjacent or turbulent trajectories in the fastest-rotating ejecta, rather than a region-dependent
effect specific to R6's geometry. The net effect is
a mild, quantifiable underestimate of R6's ejected mass and heavy-element yield relative to the
other four models (Sections~\ref{sec:massA}--\ref{sec:radioactive}), rather than a qualitative bias in
the trends reported in this paper.

\bsp
\label{lastpage}
\end{document}